\documentclass[11pt,aps,showpacs]{revtex4}

\usepackage{amsmath,amssymb,amsfonts,graphicx}

\begin{document}


\begin{titlepage}

\vskip 0.5in

\rightline{HD-THEP-03-54}

\vskip -.5in

   \title{\bf\LARGE \hskip 0.cm
      Photon mass generation during inflation:  \\
                                de Sitter invariant case \hskip 1.cm
          }

    \author{\Large Tomislav Prokopec}
\email{T.Prokopec@thphys.uni-heidelberg.de}
    \author{\Large Ewald Puchwein}
\email{E.Puchwein@thphys.uni-heidelberg.de}

\affiliation{\large Institut f\"ur Theoretische Physik, Heidelberg University,
    Philosophenweg 16, D-69120 Heidelberg, Germany}

    \date{\today}

\begin{abstract}

 We calculate the one-loop vacuum polarization tensor of scalar
electrodynamics in a locally de Sitter space-time, endowed with a nearly
minimally coupled, light scalar field. We show that the photon dynamics
is well approximated by a (local) Proca Lagrangean. Since the photon mass
can be much larger than the Hubble parameter, the photons may propagate
slowly during inflation. Finally, we briefly discuss magnetic field generation
on cosmological scales, and point out that, while the spectrum of the magnetic
field is identical to that obtained from the massless scalar,
$B_\ell \simeq B_0/\ell$, the amplitude $B_0$ may be significantly enhanced,
implying that the seed field bound for the galactic dynamo can be easily met.

\end{abstract}


\vskip 2in

\maketitle

\end{titlepage}

\tableofcontents

\section{Introduction}
\label{Introduction}

 In 1962 Julian Schwinger showed~\cite{Schwinger:1962} that
the photon of the 1+1 dimensional Quantum Electrodynamics (QED)
acquires at one loop a mass $m_\gamma = e/\sqrt{\pi}$ as a consequence
of the singular infrared behavior of the fermionic propagator.
A little later, Anderson~\cite{Anderson:1963} showed that photons
in conductive media may propagate as massive excitations. 
On the other hand, the gap generation in
superconductors has as a consequence the Meissner effect, according to which
static magnetic fields are exponentially suppressed within
superconductors. Phenomenologically, the gap can be
described by the scalar field condensate of the Landau-Ginzburg model, 
where the role of the scalar is taken by Cooper pairs. 
Abelian and non-Abelian gauge theories in thermal equilibrium
have also been studied extensively~\cite{LeBellac}. It has been found that,  
while electric fields are Debye screened, and cannot freely
propagate, magnetic fields (at the perturbative level) do
propagate as massless excitations, and screening is only dynamical. 
A realization that scalar condensates may be built by a fundamental 
scalar field, led to the establishment of the celebrated 
Higgs mechanism~\cite{EnglertBrout:1964,Higgs:1964,GuralnikHagenKibble:1964}
(sometimes referred to as the Anderson-Englert-Brout-Higgs-Kibble
mechanism), which currently represents the most popular explanation for
the origin of the mass for the electroweak gauge bosons and fermions. 
This is of course not the whole story. 
A large part of the mass of mesons and baryons
can be attributed to nonperturbative effects of a strongly coupled QCD.
For an insightful historical overview of the
development of our understanding of the concept of mass we refer
to~\cite{Okun:1991}. We paraphrase Okun: A tiny photon mass,
albeit gauge non-invariant, does not destroy the renormalizability
of QED~\cite{FeldmanMatthews:1963,ItzyksonZuber:1980}, and its
presence would not spoil the beautiful agreement between QED and
experiment. This motivates (so far unsuccessful) incessant
searches for a nonvanishing photon
mass~\cite{GoldhaberNieto:1971,FischbachKloorLangelLiuPeredo:1994}.

  Quite recently, a novel mechanism for mass generation of gauge fields has
been discovered~\cite{ProkopecTornkvistWoodard:2002prl,
ProkopecTornkvistWoodard:2002AoP,ProkopecWoodard:2003ajp,
ProkopecWoodard:2003photons}, which is operative in the presence
of rapidly (superluminally) expanding background space-times. The
concrete model within which the mechanism has been studied is
scalar electrodynamics ($\Phi$QED), endowed with a massless,
minimally coupled, scalar field evolving in a locally de Sitter
space-time background. The photon mass is radiatively induced (at
one loop) as a consequence of the canonical photon coupling to the
amplified charged scalar fluctuations, which is in contrast to the
Higgs mechanism, where the mass is induced by scalar condensate.
We now compare this gravity induced mechanism with the case of
thermal media. While magnetic fields get initially suppressed
(with respect to the conformal vacuum), at asymptotically late
times~\cite{ProkopecWoodard:2003photons}, just like in thermal
equilibrium, they are {\it not} screened, and the final amplitude
remains comparable to the conformal vacuum, $\vec B\sim \vec
B_{\rm vac}\propto a^{-2}$, where $a$ denotes the scale factor. 
On the other hand, the photon mass in
a rapidly expanding Universe implies exponentially enhanced ({\it
anti-screened}\/) electric fields,
 $\vec E \propto a^{-3/2}$, which is in contrast to the
Debye-screened electric fields of thermal media.

 Here we reconsider the mass generation mechanism of
Refs.~\cite{ProkopecTornkvistWoodard:2002prl,
ProkopecTornkvistWoodard:2002AoP,ProkopecWoodard:2003ajp,
ProkopecWoodard:2003photons} in the context of
massive scalar electrodynamics, whose Lagrangean density
in a general metric field $g_{\mu\nu}$ reads,
\begin{eqnarray}
{\cal L}_{\Phi\rm QED}  =
                       -  \frac 14\sqrt{-g} g^{\mu\rho}g^{\nu\sigma}
                                   F_{\mu\nu}F_{\rho\sigma}
                       - \sqrt{-g}g^{\mu\nu}
                            (D_{\mu} \phi)^\dagger
                            D_{\nu} \phi
                        - \sqrt{-g}(m_\phi^2 + \xi R) \phi^* \phi
\,,
\label{PhiQED}
\end{eqnarray}
where $F_{\mu\nu} \equiv \partial_{\mu} A_{\nu} - \partial_{\nu}
A_{\mu}$ is the field strength, $D_\mu = \partial_{\mu} + i e
A_{\mu}$ the covariant derivative, $g = {\rm det}[g_{\mu\nu}]$,
and $R$ is the curvature scalar. Of course, through the
Coleman-Weinberg mechanism~\cite{ColemanWeinberg:1973}, the
effective $\Phi$QED acquires radiative corrections, which can be
expressed through a renormalized mass $m_\phi\rightarrow m_R$ and
a renormalized quartic coupling, $\lambda_R$. Since we shall not
study the consequences of radiatively induced symmetry breaking in
the scalar sector, for our purposes it suffices to consider the
physical effects of a nonvanishing scalar mass, $m_\phi^2>0$. We
do however discuss briefly what happens in the case when $m_\phi^2
< 0$ (or, more precisely, when $m_\phi^2 + \xi R<0$).

 When recast in the locally de Sitter background,
\begin{equation}
  g_{\mu\nu} = a^2(\eta)\eta_{\mu\nu}
   \,,\qquad
         \eta_{\mu\nu} = {\rm diag} (-1,1,1,1)
\,,
\label{deSitter:metric}
\end{equation}
where $a$ denotes the scale factor, which -- when written in terms of
conformal time $\eta$ -- reads
\begin{equation}
        a(\eta) = - \frac{1}{H\eta}
\qquad  (\eta < 0)
\,,
\label{a(eta)}
\end{equation}
${\cal L}_{\Phi QED}$ in Eq.~(\ref{PhiQED}) reduces to,
\begin{eqnarray}
{\cal L}_{\Phi QED}   \longrightarrow
                        -  \frac 14a^{D-4} \eta^{\mu\rho}\eta^{\nu\sigma}
                                         F_{\mu\nu}F_{\rho\sigma}
                        -   a^{D-2}  \eta^{\mu\nu}
                                (\partial_{\mu} - ie A_{\mu}) \phi^*
                                (\partial_{\nu} + i e A_{\nu} ) \phi
                        -   a^D (m_\phi^2 + \xi R) \phi^* \phi
\,,
\label{PhiQED:conformal}
\end{eqnarray}
where $H$ is the Hubble parameter in inflation, 
$\partial_\mu \equiv (\partial_\eta,\nabla)$,
$D$ denotes the dimensionality of space-time, and we made use
of $\sqrt{-g} \rightarrow a^D$. Recall that
in D-dimensional de Sitter space-time, $R = D(D-1) H^2$.
In this paper we shall assume a nearly minimally coupled,
light scalar, for which $\xi \ll 1$ and $m_\phi^2 \ll H^2$. From
Eq.~(\ref{PhiQED:conformal}) it immediately follows that,
in $D=4$, gauge fields are conformally invariant, while conformal invariance
of scalar fields is granted for $\xi = 1/6$.

In section~\ref{Scalar propagator} we introduce the scalar
two-point function $G=G(\bar y)$, which (in $D$-dimensions
and in the absence of electromagnetic fields) can be expressed
in terms of the de Sitter invariant length, $\bar y$,
as the following hypergeometric function,
\begin{equation}
    G(\bar y)=\frac{\Gamma(\frac{D-1}{2}+\nu)
    \Gamma(\frac{D-1}{2}-\nu)}{(4\pi)^{{D}/{2}}
    \Gamma({D}/{2})}H^{D-2}\;
 {}_2F_1\Big(
         \frac{D-1}{2}+\nu,\frac{D-1}{2}-\nu,\frac{D}{2};1-\frac{\bar y}{4}
        \Big),
\label{G-y}
\end{equation}
where $\nu$ is given in Eq.~(\ref{eq - nu}), $\bar y = \bar
y(x;x')$ can be expressed in terms of the conformal coordinate
interval, ${\Delta \bar x}^2(x;x') 
\equiv \eta^{\mu\nu}(x_\mu-x_\mu^\prime)(x_\nu-x_\nu^\prime)
= -(\eta-\eta')^2 + \| \vec x -
\vec x\,'\|^2$, and the geodesic distance $\ell = \ell(x;x')$,
\begin{equation}
\bar y(x;x') \equiv a a' H^2 {\Delta \bar x}^2(x;x')
           = 4 \sin^2\left(\frac12 H \ell(x;x') \right)
\, .
\label{ell:deSitter}
\end{equation}
where $a\equiv a(\eta), a'\equiv a(\eta')$.

 It is instructive to expand the solution~(\ref{G-y}) around the minimally
coupled case, such that in $D=4$ we get,
\begin{equation}
 G(\bar y) \; \stackrel{D\rightarrow 4}{\longrightarrow} \;
     \frac{H^2}{4\pi^2}\bigg\{
                              \frac{1}{\bar y}
                            - \frac 12 \ln(\bar y)
                            + \frac 1{2{\tt s}}
                            - 1 + \ln(2)
                            + O({\tt s})
                    \bigg\}
\,.
\label{iDelta:4dim:massive}
\end{equation}
where the parameter $|{\tt s}|\ll 1$ is defined by,
\begin{eqnarray}
      {\tt s} &\equiv& \frac{D-1}{2}
                -    \Big[
                       \Big(
                          \frac{D-1}{2}
                       \Big)^2
                         -  \frac{m_\phi^2 + \xi R}{H^2}
                      \Big]^\frac12
                =   \frac{m_\phi^2 + \xi R}{(D-1)H^2}
                 +    O\Big(\big[(m_\phi^2 + \xi R)/H^2\big]^2\Big)
\,.
\label{s:def}
\end{eqnarray}
This is to be compared with the corresponding expression for
the massless two-point function in $D=4$,
\begin{equation}
 i \Delta(x;x')_{m =  0} \;\; \stackrel{D\rightarrow 4}{\longrightarrow} \;\;
     \frac{H^2}{4\pi^2}\bigg\{
                              \frac{\eta\eta'}{\Delta \bar x^2}
                            - \frac 12 \ln(H^2\Delta \bar x^2)
                            -\frac 14 + \ln(2)
                    \bigg\}
\,.
\label{iDelta:4dim:massless}
\end{equation}
Apart from a constant term, the two solutions~(\ref{iDelta:4dim:massive})
and~(\ref{iDelta:4dim:massless}) differ by the term
$[H^2/(8\pi^2)]\ln(aa')$ which is responsible for breakdown of
the de Sitter invariance by the massless
propagator~\cite{Allen:1985,AllenFolacci:1987,TsamisWoodard:1993,
OnemliWoodard:2002,ProkopecTornkvistWoodard:2002AoP,
ProkopecWoodard:2003photons,ProkopecWoodard:2003fermions},
and it is a consequence of the growth of scalar fluctuations during
inflation~\cite{FordParker:1977,VilenkinFord:1982}.
Note that, in the limit when ${\tt s} \rightarrow 0$,
the propagator~(\ref{iDelta:4dim:massive}) becomes formally singular,
and hence ill defined. This is a consequence of the fact that
it is {\it not} possible to construct a nontrivial de Sitter invariant
propagator for a massless scalar~\cite{Allen:1985,AllenFolacci:1987}.
Note also that, since the non-Hadamard terms in
the propagators~(\ref{iDelta:4dim:massive}) and~(\ref{iDelta:4dim:massless}) 
dominate in the infrared, they play a similar role in our mass generation 
mechanism as the fermionic states in the Schwinger mechanism, 
whose infrared behavior is sufficiently singular to lead to
mass generation in $1+1$ dimensions.

 In order to investigate how is the mass generation mechanism affected
by the difference in the scalar propagators, in
section~\ref{One-loop vacuum polarization in a locally de Sitter space}
we outline our calculation of the one-loop, renormalized, vacuum polarization
tensor of $\Phi$QED to order ${\tt s}^0$. We then use standard techniques
to arrive at the retarded tensor, which is then used in
section~\ref{Effective field equation and photon mass} to study the photon
dynamics.

 Our main result is presented in
section~\ref{Effective field equation and photon mass}, and can be
summarized as follows. To order ${\tt s}^0$, the photon dynamics
during inflation is governed by the Proca Lagrangean, which in D=4 reads,
\begin{equation}
 {\cal L}_{\rm Proca} = - \frac 14 \eta^{\mu\rho}\eta^{\nu\sigma}
                                         F_{\mu\nu}F_{\rho\sigma}
                        - \frac 12 a^2 m_\gamma^2 \eta^{\mu\nu} A_{\mu} A_{\nu}
                        + O({\tt s}^0)
\label{Proca:L}
\end{equation}
with the photon mass-squared (accurate to order ${\tt s}^0$),
\begin{equation}
    m_{\gamma}^2 \simeq \frac{\alpha H^2}{\pi {\tt s}}
                 = \frac{3\alpha H^4}{\pi(m_\phi^2+ \xi R)}
\,,
\label{photon:mass}
\end{equation}
where $\alpha = e^2/4\pi$ is the fine structure constant.
The Proca Lagrangean is derived in a generalized Lorentz gauge,
\begin{equation}
 \partial_\mu\big(a^2\eta^{\mu\nu} A_\nu\big) = 0
\,.
\label{Lorentz gauge:generalisation}
\end{equation}
This completely fixes the gauge, such that the photon contains two
transverse and one longitudinal degree of freedom. In analogy to
the Higgs mechanism, the longitudinal degree is supplied by
the charged scalar fluctuations.

 A consequence of the singular behavior of the two-point
function~(\ref{iDelta:4dim:massive}) in the limit when ${\tt s} \rightarrow 0$
is an ill-defined photon mass~(\ref{photon:mass}). The correct treatment
of the problem~\cite{ProkopecTornkvistWoodard:2002prl,
ProkopecTornkvistWoodard:2002AoP} is then to use the massless
propagator~(\ref{iDelta:4dim:massless}), in which case
one recovers the (finite) result of Ref.~\cite{ProkopecWoodard:2003photons},
which can be interpreted as the (electric) photon mass, 
$m_\gamma^2 = (2\alpha/\pi)H^2\ln(a)$. The time dependence is a consequence 
of the broken de Sitter invariance by the massless scalar 
propagator~(\ref{iDelta:4dim:massless}), which is $O(4)$ invariant. 
We expect that we would get a similar
behaviour for a light nearly minimally coupled scalar, were we using
a scalar propagator that is invariant only under the subgroup
$O(4)$ of the de Sitter group $SO(4,1)$. The divergent behaviour
of the photon mass found in Ref.~\cite{ProkopecWoodard:2003photons}
in the limit when $a\rightarrow \infty$ is related to
the $m_\phi^2 +\xi {\cal R} \rightarrow 0$ behaviour of 
the photon mass~(\ref{photon:mass}) in the de Sitter invariant case. 

Furthermore, it is interesting to compare the cases when ${\tt s}$ is small
but positive, with the case when ${\tt s}$ is small but negative.
In the case $1\gg {\tt s}>0$, the photon acquires a mass~(\ref{photon:mass}),
and the photon amplitude oscillates and gets suppressed during inflation.
On the other hand, when $-1 \ll {\tt s} < 0$,
the mass-squared~(\ref{photon:mass}) becomes negative, indicating instability
associated with a growth of the photon wave function. Consequently,
when compared with the conformal vacuum,
both electric and magnetic field can grow during inflation.
This instability may imply primordial magnetic field generation of
large amplitudes and on cosmological scales. The quantitative details
of this mechanism will be addressed elsewhere. Another crucial difference
between the mass generation in the case of a minimally coupled scalar
and that of a near minimally coupled, light scalar is in the following.
While in both cases the electric field is enhanced (anti-screened) with
respect to the conformal vacuum, the magnetic field dynamics is strongly
affected only in the latter case, in which the field is screened as
$\vec B\propto a^{-\frac 52 + \frac 12 \sqrt{1-(m_\gamma/H)^2}}$.

 Another noteworthy consequence of the photon mass generation is implied
by considering the photon dispersion relation, which can be extracted
from the Proca equation (obtained by the variation of Eq.~(\ref{Proca:L})),
\begin{equation}
  \eta^{\nu\rho}\partial_{\rho}(\partial_{\nu}A_{\mu}-\partial_{\mu}A_{\nu})
           -a^2 m_{\gamma}^2 A_{\mu}=0
\,.
\label{Proca:eom}
\end{equation}
Assuming a dependence $e^{i\vec k\cdot\vec x}$ on the spatial coordinates
and decomposing the photon field into transverse and longitudinal components,
one obtains different equations of motion for these modes. However,
in adiabatic limit, discussed  in section~\ref{sec - speed of light},
which is valid for non-relativistic modes
($k_{\rm ph} \equiv \|\vec k\|/a \ll m_\gamma$)
if $m_\gamma\gg H$ and for relativistic modes ($k_{\rm ph}\gg m_\gamma$)
if $k_{\rm ph}\gg H$ (for longitudinal modes)
or $k_{\rm ph}\gg (Hm_\gamma)^{1/2}$ (for transverse modes),
one obtains the same dispersion relation
\begin{equation}
      \omega  \simeq \sqrt{\vec k^2 + a^2 m_\gamma^2}
\end{equation}
for transverse and longitudinal modes. The propagation speed of massive photons is then given by the group velocity
\begin{equation}
   \vec v_{\rm group} = \frac{d\omega_{\rm ph}}{d\vec{k}_{\rm ph}}
                 \simeq \frac{\vec{k}_{\rm ph}}{\omega_{\rm ph}}
\,,
\end{equation}
which, for $\|\vec k_{\rm ph}\| \ll m_\gamma$ can be $\ll 1$
(recall that in our units, the speed of light in {\it vacuo}, $c=1$).
Here we used the standard definitions for the physical frequency,
$\omega_{\rm ph} \simeq \omega/a$, and the physical wave
vector, $\vec{k}_{\rm ph} = \vec k/a $.
Since, in the case when $0 < {\tt s} \ll 1$, $m_\gamma$ can be
$\gg H$, a large class of both sub- and superhorizon photon modes
may propagate with speeds, which are much smaller than the speed of
light in {\it vacuo}. We have thus discovered that, provided inflation
lasts a sufficiently long time, such that the de Sitter invariant solution for
the scalar two-point function has had enough time to get established,
light can propagate very slowly during inflation.
Moreover, because the physical momentum scales as
$\vec k_{\rm ph}\propto \vec k/a = \vec k\, {\rm e}^{-N}$, where
$N=Ht$ denotes the number of e-foldings, $t$ the physical (cosmological) time,
and $\vec k $ the conformal momentum (the physical momentum at time $t=0$,
or $\eta = -1/H$),
the group velocity at asymptotically late times drops exponentially with
time, $v_{\rm group}\propto {\rm e}^{-Ht}$
($t\rightarrow \infty$).
We also note that, if the adiabatic conditions are not met,
one can still calculate ${\vec v}_{\rm group} = d\omega/d\vec k$
by solving the equation for $\omega = \omega(\vec k,\eta)$ exactly.
As an example, in section~\ref{sec - speed of light} we show that
the relativistic longitudinal photons, for which
$ k_{\rm ph}, H \gg m_\gamma$, propagate superluminally.

The mass-induced slow-down of light differs significantly from
the light slow-down reported not a long time
ago~\cite{HauHarrisDuttonBehroozi:1999}
in ultra low temperature sodium gas, where the slow-down is induced by
a steep frequency dependence of the index of refraction. Since the effect is
of a resonant origin, it pertains only in a very narrow range of frequencies,
which is in contrast to the mass-induced slow-down during inflation,
which is effective for a broad range of frequencies.
We also mention the work of Ref.~\cite{LeonhardtPiwnicki:1999},
where an extreme version of the Fresnel effect is
considered, in which superfluid vortices moving at a superluminal speed
are used to generate an optical Aharonov-Bohm effect. Moreover,
if they rotate faster than the speed of light in the medium,
they can trap the slow light, and in this sense mimic the
black holes of general relativity.

Finally, in section~\ref{Physical implications of our results} we
present an estimate of the cosmological magnetic fields produced
as a consequence of a radiatively induced photon mass during inflation.
Our conclusion is that the field strengths thus produced are sufficiently
strong to satisfy the bounds on the seed magnetic field
of the galactic dynamo mechanism.
Appendices are reserved for technical details of the calculations.

\section{Scalar propagator}
\label{Scalar propagator}

A complex scalar field of scalar electrodynamics~(\ref{PhiQED:conformal})
obeys (in the absence of electromagnetic fields) the following equation
of motion,
\begin{equation}
    \eta^{\mu\nu}\partial_{\mu}a^{D-2}\partial_{\nu}\phi
                   - a^D(m_\phi^2+\xi R)\phi=0.
\label{Phi:eom}
\end{equation}
Of course, the two-point Wightman functions,
$\langle \alpha|\phi(x)\phi^{\dagger}(x')|\alpha\rangle$ and
$\langle \alpha|\phi^{\dagger}(x')\phi(x)|\alpha\rangle$,
satisfy the same differential equation.
For a de Sitter invariant vacuum state
$|\alpha\rangle$, where $\alpha$ represents the (real)
parameter that classifies all globally de Sitter invariant
vacua~\cite{Allen:1985}
({\it cf.} also Refs.~\cite{SchomblondSpindel:1976} and~\cite{Mottola:1985}),
they can depend only on the geodesic distance between $x$
and $x'$, and may be written as a function of
$z(x,x') \equiv 1- {\bar y(x,x')}/{4}$ only
({\it cf.} Eq.~(\ref{ell:deSitter})).
For such a function $\hat G(z)$ the differential equation can be recast as,
\begin{equation}
     \frac{\eta^{\mu\nu}(\partial_{\mu}z)(\partial_{\nu}z)}{a^2}
                 \frac{d^2}{dz^2}\hat G
      +  \Big[
              \frac{\eta^{\mu\nu}\partial_{\mu}\partial_{\nu}z}{a^2}
           -  \frac{(D-2)H(\partial_0 z)}{a}
         \Big]\frac{d}{dz}\hat G
      -  (m_\phi^2+\xi R)\hat G
      = 0
\,.
\end{equation}
Upon calculating the derivatives of $z$, and contracting with $\eta^{\mu\nu}$,
one obtains the hypergeometric differential equation,
\begin{equation}
         z(1-z)\frac{d^2}{dz^2}\hat G
       + D\Big(\frac{1}{2}-z\Big)\frac{d}{dz}\hat G
       - (m_\phi^2+\xi R)\hat G
       = 0
\,.
\label{eq - Prop(z) eom}
\end{equation}
Taking account of the $z \Leftrightarrow 1-z$ symmetry of
this equation, one can write the general solution for
$G(\bar y)\equiv \hat G(z)$ in terms of the hypergeometric
functions~\cite{ChernikovTagirov:1968,Tagirov:1972,BunchDavies:1978},
\begin{eqnarray}
        G(\bar y)= c \phantom{\;}_2F_1\Big(\frac{D-1}{2} + \nu,
                                          \frac{D-1}{2}-\nu,
                                          \frac{D}{2};
                                          1-\frac{\bar y}{4}
                                      \Big)
           +c^\prime \phantom{\,}_2F_1\Big(\frac{D-1}{2}+\nu,
                                           \frac{D-1}{2}-\nu,
                                           \frac{D}{2};
                                           \frac{\bar y}{4}\Big)
\,,
\label{eq - Prop general solution}
\end{eqnarray}
where (so far) $c$ and $c^\prime$ are arbitrary constants
(which are dependent on $\alpha$), and
\begin{equation}
    \nu=\bigg[{\left(\frac{D-1}{2}\right)^2
           -  \frac{m_\phi^2 + \xi R}{H^2}}
        \bigg]^\frac 12
\,.
\label{eq - nu}
\end{equation}

The hypergeometric function becomes singular~\cite{Allen:1985}, when the last
argument approaches 1 or $-1$, and has a branch cut from 1 ($-1$) to $+\infty$
($-\infty$). In order to completely specify the Wightman functions, we demand
that there be a singularity only if $x$ and $x'$ are light-like related.
Furthermore, at short distances this singularity should have the Hadamard form
of the Minkowski space two-point functions,
since (at short distances) the scalar field can be
only weakly affected by the expansion of space-time.
With these assumptions one finds
$c^\prime = 0$, such that~(\ref{eq - Prop general solution}) reduces to
\begin{equation}
    G(\bar y) = \frac{\Gamma(\frac{D-1}{2}+\nu)
             \Gamma(\frac{D-1}{2}-\nu)}{(4\pi)^{\frac{D}{2}}
              \Gamma(\frac{D}{2})}H^{D-2}
             \phantom{\;}_2F_1\Big(
                                   \frac{D-1}{2}+\nu,
                                   \frac{D-1}{2}-\nu,
                                   \frac{D}{2};
                                   1-\frac{\bar y}{4}
                              \Big)
\label{the propagator}
\,.
\end{equation}
where the precise value of $c$ is dictated by the canonical
commutation relation of $\phi$ and its canonical
momentum~\cite{SchomblondSpindel:1976}. This choice corresponds to
the unique vacuum, $|\alpha = -\infty\rangle$. (The $\alpha = -\infty $
vacuum is in literature known under various names:
Chernikov-Tagirov vacuum, Bunch-Davies vacuum, Euclidean vacuum,
thermal vacuum.) The vacua with $\alpha\neq -\infty$ contain an additional
singularity at the antipodal point, $\bar y = 4$ (which
corresponds to $\eta^\prime = -\eta$ and $\vec x^{\,\prime} = \vec
x$), which can be also interpreted as a constant
($\alpha$-dependent) nonvanishing particle number at all momenta,
such that $\alpha\neq -\infty$ vacua are associated to a divergent
stress-energy tensor. While this divergence should be regularized, it
can be done so only for a particular choice of $\alpha$. In our
point of view, the requirement that, at short distances, one
should recover the Hadamard form for the singularity of two-point
functions (which, by the way, has been tested in particle
accelerators), singles out the $\alpha = -\infty $ vacuum in a natural
way. Indeed, the Hadamard form can be violated only at the price
of giving up locality in position space. The nonlocality is namely
necessary if the Hubble scale physics is to influence the physics
at much shorter scales, responsible for the Hadamard singularity.

Analogous to different pole prescriptions in the momentum space
integrals for the various forms of the scalar
propagator~\cite{PeskinSchroeder:1995}, one can write the
(anti-)Feynman and Wightman propagators as the same function G
of the appropriately modified de Sitter length functions,
\begin{equation}
    i\Delta_{bb'}(x,x')=G(y_{bb'})
\qquad (b,b' = +, -)
\,,
\label{eq - prop +-+-}
\end{equation}
where
\begin{align}
    i\Delta(x,x')\equiv
    i\Delta_{++}(x,x')&\equiv \langle 0|
                                        T[\phi(x)\phi^{\dagger}(x')]
                              |0\rangle
\qquad
 ({\tt Feynman})
\label{eq - prop start}
\\
    i\Delta_{+-}(x,x')&\equiv \langle 0|
                                        \phi^{\dagger}(x')\phi(x)
                              |0\rangle
\qquad\quad\,\,
 ({\tt Wightman})
\\
    i\Delta_{-+}(x,x')&\equiv \langle 0|
                                        \phi(x)\phi^{\dagger}(x')
                               |0\rangle
\qquad\quad\,\,
 ({\tt Wightman})
\\
    i\Delta_{--}(x,x')&\equiv \langle 0|
                                        \bar T[\phi(x)\phi^{\dagger}(x')]
                               |0\rangle
\qquad
 ({\tt anti\!\!-\!\!Feynman})
\,.
\label{eq - prop stop}
\end{align}
Here $T$ ($\bar T$) denotes time (anti-time) ordering and
\begin{equation}
    y_{bb'}=\frac{\Delta x_{bb'}^2}{\eta\eta'}
\,,
\label{eq - y +-+-}
\end{equation}
with
\begin{align}
    \Delta x^2\equiv
    \Delta
    x_{++}^2=&-(|\eta-\eta'|-i\delta)^2+\|\vec{x}-\vec{x}'\|^2,
\label{eq - def dx2}\\
    \Delta
    x_{+-}^2=&-(\eta-\eta'+i\delta)^2+\|\vec{x}-\vec{x}'\|^2,
\label{eq - def dx2+-}\\
    \Delta
    x_{-+}^2=&-(\eta-\eta'-i\delta)^2+\|\vec{x}-\vec{x}'\|^2,
\label{eq - def dx2-+}\\
    \Delta
    x_{--}^2=&-(|\eta-\eta'|+i\delta)^2+\|\vec{x}-\vec{x}'\|^2.
\end{align}
 From now on we shall use the notation,
\begin{equation}
    y\equiv y_{++}=\frac{\Delta x^2}{\eta\eta'}
\,.
\label{eq - def y}
\end{equation}

In order to make progress toward calculating the vacuum polarization
tensor in dimensional regularization, we shall expand the scalar
propagator~(\ref{the propagator})--(\ref{eq - prop stop})
around the massless minimally coupled case and in the vicinity of $D=4$.
To this purpose, it is useful to define the parameter {\tt s} as,
\begin{equation}
    {\tt s} \equiv \frac{D-1}{2}-\nu,
\label{eq - def s}
\end{equation}
where $\nu$ is defined in~(\ref{eq - nu}), and the parameter
\begin{equation}
  \varepsilon \equiv  4 - D
\,.
\label{varepsilon}
\end{equation}
For a light and/or nearly minimally coupled scalar, for which  $m_\phi \ll H$
and $|\xi | \ll 1$, we can expand $\nu$ in powers of
$({m_\phi^2+ \xi R})/{H^2}$,
to obtain the following approximation for {\tt s},
\begin{equation}
    {\tt s} = \frac{1}{D-1}\frac{m_\phi^2+ \xi R}{H^2}
            + O\left(\Big(\frac{m_\phi^2+ \xi R}{H^2}\Big)^2\right)
\,.
\end{equation}
In Appendix~\ref{sec - app prop expansion} we show
(see Eq.~(\ref{eq - app expanded prop})) that, when expanded
in {\tt s}, $y$ and $\varepsilon$, the Feynman
propagator~(\ref{eq - prop +-+-}--\ref{eq - prop start}) reduces to,
\begin{equation}
    i\Delta(x,x') = \beta f(y)
                  + {\tt s}\frac{H^{2-\varepsilon}}
                                {(4\pi)^{2-\frac{\varepsilon}{2}}}
                                \big[g(y)+h(y)\big]
                  + O(\varepsilon,{\tt s}^2)
\,,
\label{eq - prop minimally}
\end{equation}
where
\begin{eqnarray}
 \beta &=& \Big(\frac{H}{2\pi}\Big)^2
            \Big(\frac{\pi}{H^2}\Big)^{\frac{\varepsilon}{2}}
            \Gamma\Big(2-\frac{\varepsilon}{2}\Big),
\label{eq - def beta}
\\
   f(y) &=& \frac{1}{1-\frac{\varepsilon}{2}}\,
           \frac{1}{y^{1-\frac{\varepsilon}{2}}}
         - \Big(1-\frac{\varepsilon}{4}\Big)
           \frac{y^{\frac{\varepsilon}{2}}}{\varepsilon}
         + \frac{2^{\varepsilon}\Gamma(3-\varepsilon)}
                {4\Gamma^2(2-\frac{\varepsilon}{2})}
         \Big[
              \frac{1}{{\tt s}}
            + \pi\cot\Big(\frac{\pi\varepsilon}{2}\Big)
            - \gamma_{\rm E}
            - \psi(3-\varepsilon)
         \Big],
\label{eq - def f}
\\
    g(y) &=& \frac{\Gamma(3-\varepsilon)}
                  {\Gamma(2-\frac{\varepsilon}{2})}
             \frac{C({\tt s},\varepsilon)}{{\tt s}}
          - \Big(\frac{y}{4}\Big)^{\frac{\varepsilon}{2}}
                 \frac{\Gamma(3-\frac{\varepsilon}{2})}
                      {\frac{\varepsilon}{2}}
            \Big[
                 \pi\cot\Big(\frac{\pi\varepsilon}{2}\Big)
               - \psi\Big(3-\frac{\varepsilon}{2}\Big)
               + \psi\Big(\frac{\varepsilon}{2}\Big)
            \Big],
\label{eq - def g}
\\
    h(y) &=& \sum_{n=1}^{\infty}\frac{2-n(n+2)
             \ln({y}/{4})}{n^2}
             \Big(\frac{y}{4}\Big)^{n}
\,.
\label{eq - def h}
\end{eqnarray}
Here $\Gamma$ denotes the Euler-Gamma function,
$\gamma_{\rm E}\approx 0.577$ the Euler-Mascheroni constant, $\psi$
is defined by $\psi(z) \equiv \Gamma'(z)/\Gamma(z)$ and
$C({\tt s},\varepsilon)$ is an order {\tt s} term that is independent on the
coordinates. We will later find that it does not contribute to the
vacuum polarization at order ${\tt s}^0$. The $O(\varepsilon)$ term
in the infinite sum is not important, because it will not be needed for the
dimensional regularization and renormalization of the vacuum polarization
tensor outlined in
section~\ref{One-loop vacuum polarization in a locally de Sitter space}
and Appendix~\ref{sec - vac pol}.

For calculating the one-loop vacuum polarization tensor in a locally
de Sitter space-time, we will also need the coincident limit
$x'\rightarrow x$ (equivalently $y \rightarrow 0$) of the
propagator~(\ref{eq - prop minimally}). To obtain this limit in
dimensional regularization, one must assume that $\varepsilon$ is
large enough that a vanishing quantity like $y$ is raised to non-negative
powers only. Thus we get
\begin{equation}
  \lim_{x'\rightarrow x}i\Delta(x,x')
                 = \beta\frac{2^{\varepsilon}\Gamma(3-\varepsilon)}
                              {4\Gamma^2(2-\frac{\varepsilon}{2})}
                   \Big[\frac{1}{{\tt s}}
                      + \pi\cot\Big(\frac{\pi\varepsilon}{2}\Big)
                      - \gamma_{\rm E}
                      - \psi(3-\varepsilon)
                   \Big]
                 + O({\tt s})
\,.
    \label{eq - lim prop}
\end{equation}
This is in contrast with the massless propagator~(\ref{iDelta:4dim:massless}),
which grows logarithmically with the scale factor during inflation,
$i\Delta(x,x)_{m\rightarrow 0} \propto \ln(a)$.

\section{One-loop vacuum polarization in a locally de Sitter space}
\label{One-loop vacuum polarization in a locally de Sitter space}

\begin{figure}
    \centering
    \includegraphics[scale=0.4]{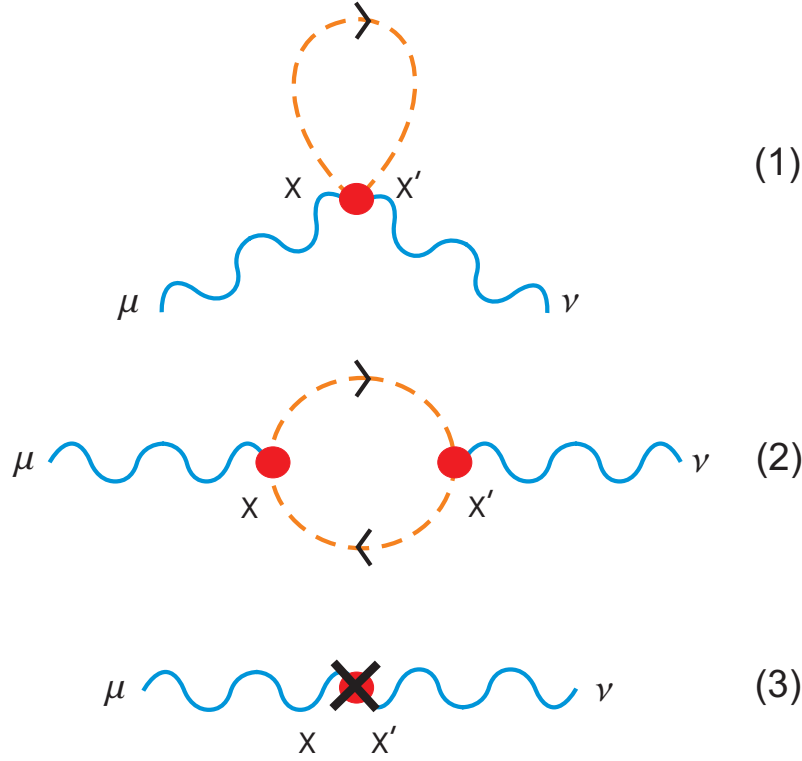}
    \caption{The one-loop Feynman graphs contributing to the vacuum
     polarization tensor in scalar QED.}
    \label{fig - vac pol graphs}
\end{figure}

 We are now ready to calculate the vacuum polarization
tensor of scalar electrodynamics~($\Phi$QED) in curved
space-time backgrounds~(\ref{PhiQED}), (\ref{PhiQED:conformal})
in the one-loop approximation,
based on which one can study the dynamics of photons during inflation
({\it cf.} Ref.~\cite{ProkopecTornkvistWoodard:2002AoP}).
The relevant diagrams contributing
to the one-loop polarization tensor are shown in
figure~\ref{fig - vac pol graphs}.
The counterterm in figure~\ref{fig - vac pol graphs}.(3) is added in order to
cancel the divergence, appearing in dimensional regularization
in the limit when $D\rightarrow 4$.
Using the position-space Feynman rules of
Appendix~\ref{sec - feynman rules sQED}, one finds the following expressions
for the one-loop graphs shown in figure~\ref{fig - vac pol graphs}
\begin{align}
 i[\phantom{\,}^\mu\Pi^{\nu}]^{(1)}(x,x')
       = & -2ie^2\sqrt{-g(x)}g^{\mu\nu}(x)i\Delta(x,x)\delta^D(x-x'),\\
       = & -2ie^2a^{D-2}\eta^{\mu\nu}i\Delta(x,x)\delta^D(x-x')
    \label{eq - fig.1}
\end{align}
\begin{align}
  i[\phantom{\,}^\mu\Pi^{\nu}]^{(2)}(x,x')
       = 2 & e^2\sqrt{-g(x)}g^{\mu\rho}(x)\sqrt{-g(x')}g^{\nu\sigma}(x')
\nonumber\\
       \times &
           \big[(\partial_{\rho}i\Delta(x,x'))\partial'_{\sigma}i\Delta(x,x')
         - i\Delta(x,x')\partial_{\rho}\partial'_{\sigma}i\Delta(x,x')\big]
\\
       = 2 &e^2a^{D-2}a'^{D-2}\eta^{\mu\rho}\eta^{\nu\sigma}
         \big[(\partial_{\rho}i\Delta(x,x'))(\partial'_{\sigma}i\Delta(x,x'))
        - i\Delta(x,x')\partial_{\rho}\partial'_{\sigma}i\Delta(x,x')\big]
,
    \label{eq - fig.2}
\end{align}
where $i\Delta(x,x')=i\Delta(x',x)$ is used and
\begin{align}
i[\phantom{\,}^\mu\Pi^{\nu}]^{(3)}(x,x')
    =&-i\delta
    Z\partial_{\rho}\big(\sqrt{-g(x)}[g^{\mu\nu}(x)g^{\rho\sigma}(x)-g^{\mu\sigma}(x)g^{\nu\rho}(x)]\partial_{\sigma}'\delta^D(x-x')\big),\\
    =&-i\delta Z[{}^\mu P^\nu]
    a^{D-4}\delta^D(x-x').
\label{eq - fig.3}
\end{align}
Here $a$ and $a'$ are defined by $a\equiv a(\eta)$, $a'\equiv a(\eta')$,
and $[{}^\mu P^\nu]$ is the transverse projector defined in
Eq.~(\ref{projector-operators}) below.
One can use Eq.~(\ref{eq - lim prop}) for $i\Delta(x,x)$ in
(\ref{eq - fig.1}) and (\ref{eq - prop minimally}) for
$i\Delta(x,x')$ in (\ref{eq - fig.2}). Then the sum of the graphs
Fig.~\ref{fig - vac pol graphs}.(1) and Fig.~\ref{fig - vac pol
graphs}.(2), $i[^{\mu}\Pi^{\nu}_{1+2}]$ can be derived, one
obtains (see Appendix~\ref{sec - vac pol})
\begin{eqnarray}
i[^{\mu}\Pi^{\nu}_{1+2}](x,x')
         &=& \frac{\alpha}{2\pi^3}
 \bigg\{
        \frac{\pi^{\varepsilon}\Gamma^2(1-\frac{\varepsilon}{2})}
             {2(3-\varepsilon)}
         \big[\phantom{\,}^\mu P^\nu\big]
        \frac{1}{\Delta x^{4-2\varepsilon}}
    -         \big[\phantom{\,}^\mu P^\nu\big]
       \bigg[
             \frac{\frac{1}{2}\ln(\frac{y}{4})-\frac{1}{2\tt s}+2}
                  {\eta\eta'\Delta x^{2}}
          +  \frac{\ln(\frac{y}{4})}{2(y-4)\eta^2\eta'^2}
       \bigg]
\nonumber\\
 &&\hskip 0.7cm
   +\;         \big[\phantom{\,}^\mu\bar P^\nu\big]
       \frac{1}{\eta^2\eta'^2}
       \bigg[
             \frac{1}{8}\ln^2\Big(\frac{y}{4}\Big)
           + \ln\Big(\frac{y}{4}\Big)\Big(1-\frac{1}{4\tt s}
                                        + \frac{y-2}{2(y-4)}\Big)
           - \frac{1}{4}Li_2\Big(1-\frac{y}{4}\Big)
       \bigg]
\nonumber\\
 &&\hskip 0.7cm
   +\;O({\tt s})
 \bigg\},
    \label{eq - ipi1+2 main}
\end{eqnarray}
where
\begin{eqnarray}
\big[\phantom{\,}^\mu P^\nu\big] &=&
      \eta^{\mu\rho}\eta^{\nu\sigma}(\eta_{\rho\sigma} \partial'\cdot\partial
                                   - \partial'_{\rho}\partial_{\sigma}
                                    )
\nonumber\\
\big[\phantom{\,}^\mu\bar P^\nu\big] &=&
      \eta^{\mu i}\eta^{\nu j}(\eta_{ij} \nabla^\prime\cdot\nabla
                                   - \partial'_{i}\partial_{j}
                                    )
\label{projector-operators}
\end{eqnarray}
are the manifestly transverse projector operators,
$\partial_\mu[\phantom{\,}^\mu P^\nu] = 0 =
\partial_\nu^\prime [\phantom{\,}^\mu P^\nu]$,
$\partial_\mu[\phantom{\,}^\mu \bar P^\nu] = 0 =
\partial_\nu^\prime [\phantom{\,}^\mu \bar P^\nu]$, and
$\partial^\prime\cdot\partial 
  \equiv \eta^{\mu\nu}\partial_\mu^\prime\partial_\nu$,
$\nabla^\prime\cdot\nabla \equiv \partial_i^\prime\partial_i$
(greek indices $\alpha, \beta, ..$ run from $0$ to $3$, 
while latin indices $i,j, ..$ run from $1$ to $3$).
In Eq.~(\ref{eq - ipi1+2 main}) D was taken to 4 in all terms,
except in the first, because only this term requires regularization.
$\Delta x^2$ and $y$ are given by (\ref{eq - def dx2}) and~(\ref{eq - def y}),
and $Li_2$ is the dilogarithm function defined by
$Li_2(z)\equiv-\int_0^{z}\frac{\ln(1-t)}{t}dt$.

 The first term in Eq.~(\ref{eq - ipi1+2 main}) is exactly
what one would find for Fig.\ref{fig - vac pol graphs}.(1) and
Fig.~\ref{fig - vac pol graphs}.(2) for a massless scalar field in
flat space~\cite{ProkopecTornkvistWoodard:2002AoP}. Its divergence for
$\varepsilon\rightarrow0$ can be seen from
\begin{equation}
    \frac{1}{\Delta
    x^{4-2\varepsilon}}=-\frac{1}{2\varepsilon(1-\varepsilon)}\partial^2\frac{1}{\Delta
    x^{2-2\varepsilon}}.
\end{equation}
Combining this with
\begin{equation}
    \partial^2\frac{1}{\Delta
    x^{2-\varepsilon}}=\frac{4i\pi^{2-\frac{\varepsilon}{2}}}{\Gamma(1-\frac{\varepsilon}{2})}\delta^D(x-x'),
\end{equation}
one finds
\begin{equation}
    \frac{1}{\Delta x^{4-2\varepsilon}}=
    -\frac{\partial^2}{2\varepsilon(1-\varepsilon)}\Big[\frac{1}{\Delta x^{2-2\varepsilon}}
    -\frac{\mu^{-\varepsilon}}{\Delta x^{2-\varepsilon}}\Big]
    -\frac{2\pi^2i(\sqrt{\pi}\mu)^{-\varepsilon}}{\varepsilon(1-\varepsilon)\Gamma(1-\frac{\varepsilon}{2})}\delta^{D}(x-x'),
    \label{eq - 1 / dx^(4-2epsilon)}
\end{equation}
where $\mu$ is a parameter that will be used for regularization.
Taking $\varepsilon\rightarrow0$ in the first term gives
\begin{equation}
    \rightarrow-\frac{\partial^2}{4}
    \Big[\frac{\ln(\mu^2\Delta x^2)}{\Delta x^2}
    \Big]
  - \frac{2\pi^2i(\sqrt{\pi}\mu)^{-\varepsilon}}
         {\varepsilon(1-\varepsilon)\Gamma(1-\frac{\varepsilon}{2})}
    \delta^{D}(x-x')
\,.
\label{extracting delta function}
\end{equation}
This expression can be used for the first term of (\ref{eq -
ipi1+2 main}). Its divergent local part can be canceled by the
counterterm Fig.\ref{fig - vac pol graphs}.(3), which can be
rewritten as
\begin{align}
    [\phantom{\,}^\mu\Pi^{\nu}]^{(3)}(x,x')
  = -i\delta Z [\phantom{\,}^\mu P^\nu]
    (1-\varepsilon\ln(a)+O(\varepsilon^2))\delta^D(x-x')
\,,
\label{eq - fig.3 rewritten}
\end{align}
where $\delta Z$ is chosen such that, to the lowest order in $\varepsilon$, the
counter-term cancels the divergence in Eq.~(\ref{extracting delta function}).
Note that the logarithm in (\ref{eq - fig.3 rewritten}) still gives a
contribution to the renormalized vacuum polarization, which
grows logarithmically with the scale factor. Thus we find
\begin{align}
    i[^{\mu}\Pi^{\nu}_{ren}]
    =&\frac{\alpha}{2\pi^{3}}
 \bigg\{
        - \big[\phantom{\,}^\mu P^\nu\big]
    \bigg[\partial^2\bigg(\frac{\ln(\mu^2 \Delta x^2)}{24 \Delta x^2}\bigg)
        + \frac{i \pi^2}{3}\ln(a)\delta^4(x-x')
    \bigg]
\nonumber \\
    &\hskip 1cm -\big[\phantom{\,}^\mu P^\nu\big]
    \bigg[
          \frac{\frac{1}{2}\ln\big(\frac{y}{4}\big)-\frac{1}{2\tt s}+2}
               {\eta\eta'\Delta x^{2}}
     + \frac{\ln\big(\frac{y}{4}\big)}{2(y-4)\eta^2\eta'^2}
    \bigg]
\nonumber\\
    &\hskip 1cm  + \big[\phantom{\,}^\mu \bar P^\nu\big]
\frac{1}{\eta^2\eta'^2}
    \bigg[
          \frac{1}{8}\ln^2\Big(\frac{y}{4}\Big)
        + \ln\Big(\frac{y}{4}\Big)\Big(
                                       1 - \frac{1}{4\tt s}
                                    + \frac{y-2}{2(y-4)}
                                  \Big)
    -\frac{1}{4}Li_2\Big(1-\frac{y}{4}\Big)
    \bigg]
\nonumber\\
     &\hskip 1cm  + O({\tt s})
 \bigg\}
\,,
\label{eq - ren vac pol}
\end{align}
which is the renormalized vacuum polarization tensor for $D=4$
written in a manifestly transverse form.

\subsection{The retarded vacuum polarization tensor}
\label{The retarded vacuum polarization tensor}

In this section we use the Schwinger-Keldysh
formalism~\cite{Schwinger:1961,Keldysh:1964,ChouSuHaoYu:1985,Jordan:1986} to
construct the retarded vacuum polarization tensor,
$\big[\phantom{\,}^\mu\Pi^{r,\nu}_{\rm ren}\big]$, required to
study the dynamics of photons during inflation. In order to do
that, one is led to modify the Feynman rules of Appendix~\ref{sec
- feynman rules sQED}, such that, in addition, the vertices become
signed as $b=+$ or $-$, while the propagators~(\ref{eq - prop
+-+-}--\ref{eq - prop stop}) (which connect these vertices)
acquire two signed indices, $i\Delta_{bb'}(x,x')\;$  ($b,b' =
+,-$). Repeating the procedure of section~\ref{One-loop vacuum
polarization in a locally de Sitter space}, in which we calculated
the Feynman vacuum polarization tensor,
$i\big[\phantom{\,}^\mu\Pi_{++}^\nu\big](x,x')$, we shall now
outline how to derive other relevant vacuum polarization tensors,
$i\big[\phantom{\,}^\mu \Pi_{bb'}^\nu\big]$. Since the different
propagators (\ref{eq - prop +-+-}) are the same function of the
appropriately modified de Sitter length functions~(\ref{eq - y
+-+-}), in order to get the different versions of the polarization
tensor, we just have to use~(\ref{eq - y +-+-}) in~(\ref{eq - ren
vac pol}). A couple of subtle remarks are in order. In the
derivation of the vacuum polarization tensor we used
$i\Delta(x,x')=i\Delta(x',x)$. While $i\Delta_{++}(x,x')$ and
$i\Delta_{--}(x,x')$ are symmetric under the exchange of $x$,
$x'$, for the `off-diagonal' Wightman functions $i\Delta_{+-}$
and $i\Delta_{-+}$, the following symmetry of the propagators
ought to be used
\begin{align}
    i\Delta_{bb'}(x,x')=i\Delta_{b'b}(x',x)
\qquad (b,b' = +,-)
\,,
\end{align}
which can be easily established from $\Delta
x^2_{bb'}(x,x')=\Delta x^2_{b'b}(x',x)$. Moreover, because the
vertices are signed, the off-diagonal vacuum polarization tensors,
$i[^{\mu}\Pi^{\nu}_{+-}]$ and $i[^{\mu}\Pi^{\nu}_{-+}]$, acquire
an overall {\it minus} sign with respect to
$i[^{\mu}\Pi^{\nu}_{++}]$. Finally, there are no $+-$ or $-+$
seagull graphs or countertems, and there are no local terms coming
from Fig.~\ref{fig - vac pol
graphs}.(2)~\cite{ProkopecTornkvistWoodard:2002AoP}. Based on the
above considerations and~Eq.~(\ref{eq - ren vac pol}), we can
write,
\begin{align}
\!\!\!\!\!
\big[^{\mu}\Pi^{\nu}_{bb'}\big]
   =&\frac{i\alpha}{2\pi^3}bb'
\bigg\{
    \big[\phantom{\,}^{\mu}P^{\nu}\big]
    \bigg[\partial^2\frac{\ln(\mu^2 \Delta x^2_{bb'})}{24 \Delta x^2_{bb'}}
    +\frac{i \pi^2}{6}(b+b')\ln(a)\delta^4(x-x')
    \bigg]
\nonumber \\
    &\hskip 0.9cm
     +   \big[\phantom{\,}^{\mu}P^{\nu}\big]
    \bigg[
    \frac{\frac{1}{2}\ln({y_{bb'}}/{4})-\frac{1}{2{\tt s}}+2}
                   {\eta\eta'\Delta x^2_{bb'}}
       +  \frac{\ln({y_{bb'}}/{4})}{2(y_{bb'}-4)\eta^2\eta'^2}
    \bigg]
\nonumber\\
    &\hskip 0.9cm
     -\big[\phantom{\,}^{\mu}\bar P^{\nu}\big] \frac{1}{\eta^2\eta'^2}
    \bigg[
       \frac{1}{8}\ln^2\Big(\frac{y_{bb'}}{4}\Big)
     +\ln\Big(\frac{y_{bb'}}{4}\Big)
          \Big(1-\frac{1}{4{\tt s}}+\frac{y_{bb'}-2}{2(y_{bb'}-4)}\Big)
    -\frac{1}{4}Li_2\Big(1-\frac{y_{bb'}}{4}\Big)
    \bigg]
\bigg\}
\nonumber\\
    &\hskip 0.9cm + O({\tt s})
\,.
\label{eq - vac pol versions}
\end{align}
For the derivation of an effective field equation of the photon
field we shall need the retarded vacuum polarization tensor, which is
defined by
\begin{equation}
    [^{\mu}\Pi^{r,\nu}_{ren}](x,x')=[^{\mu}\Pi^{\nu}_{++}](x,x')+[^{\mu}\Pi^{\nu}_{+-}](x,x').
\end{equation}
Because the two tensors contribute with an overall minus sign
({\it cf.} the sign prefactor $bb'$ in Eq.~(\ref{eq - vac pol
versions})), the contributions to the retarded vacuum polarization
come only from branch cuts and singularities of~(\ref{eq - vac pol
versions}) in $\Delta x^2_{bb'}$ and $y_{bb'}$ and from the term
$\propto (b+b')\ln(a)\delta^4(x-x')$. In order to extract these
cut and pole contributions, the following formulas are useful,
\begin{eqnarray}
      \partial^2\bigg[
                      \frac{\ln(\mu^2 \Delta x^2_{bb'})}
                           {24 \Delta x^2_{bb'}}
                \bigg]
   &=& \partial^4\bigg[
                       \frac{1}{192}\ln^2(\mu^2 \Delta x^2_{bb'})
                     - \frac{1}{96}\ln(\mu^2 \Delta x^2_{bb'})
                 \bigg]
\nonumber\\
    \frac{\frac{1}{2}\ln\big(\frac{y_{bb'}}{4}\big) - \frac{1}{2\tt s}+2}
         {\eta\eta'\Delta x^{2}_{bb'}}
   &=& \frac{1}{8\eta\eta'}
 \bigg[\frac{1}{2}\partial^2(\ln^2(\Delta x^2_{bb'}))
     + \Big(3-\ln(4\eta\eta')
          - \frac{1}{\tt s}
       \Big)\partial^2 \ln(\Delta x^2_{bb'})
 \bigg]
\,,
\end{eqnarray}
from which it can be seen that the contributions, which are singular at the
light-cone, yield finite cut and/or pole contributions.
The contributions from the logarithms in~(\ref{eq - vac pol versions})
are simply,
\begin{eqnarray}
 \ln(\Delta x^2_{++})-\ln(\Delta x^2_{+-})
  &=& \ln(\frac{y_{++}}{4})-\ln(\frac{y_{+-}}{4})
   =  2i\pi\,\Theta(\Delta\tau^2)\Theta(\Delta\eta)
\\
 \ln^2(\frac{y_{++}}{4})-\ln^2(\frac{y_{+-}}{4})
  &=& 4i\pi\ln\Big|\frac{\Delta \tau^2}{4\eta\eta^\prime}\Big|
           \,\Theta(\Delta\tau^2)\Theta(\Delta\eta)
\\
 \ln^2(\Delta x^2_{++})-\ln^2(\Delta x^2_{+-})
  &=& 4i\pi\ln|\Delta \tau^2|
            \,\Theta(\Delta\tau^2)\Theta(\Delta\eta)
\,,
\end{eqnarray}
where $\Delta \tau^2 \equiv \Delta \eta^2 - \|\Delta \vec x\|^2$,
$\Theta$ is the Heaviside step function,
$\Theta(\Delta \tau^2) = \Theta(|\Delta\eta| - \|\vec x\|)$,
$\Delta\eta\equiv\eta-\eta'$ and
$\Delta\vec{x}\equiv\vec{x}-\vec{x}'$. From the integral
representation of the dilogarithm function one finds,
\begin{equation}
    Li_2\Big(1-\frac{y_{++}}{4}\Big) - Li_2\Big(1-\frac{y_{+-}}{4}\Big)
  = - 2i\pi\,\ln\Big(1+\frac{\Delta \tau^2}{4\eta\eta^\prime}\Big)
    \,\Theta(\Delta\tau^2)\Theta(\Delta\eta).
\end{equation}
Combining these equations we find the following expression for the
renormalized, retarded vacuum polarization tensor to one-loop order,
\begin{eqnarray}
 [^{\mu}\Pi^{r,\nu}_{ren}](x,x') \!\!&=&\!\! \frac{\alpha}{2\pi^2}
\bigg\{ \!-\! \big[\phantom{\,}^{\mu} P^{\nu}\big]
    \Big\{\frac{1}{48}\partial^4
          \Big[
               \Theta(\Delta\tau^2)\Theta(\Delta\eta)
                        (\ln|\mu^2\Delta \tau^2|-1)
          \Big]
       +  \frac{\pi}{3}\ln(a)\delta^4(x-x')
    \Big\}
\nonumber\\
    &-& \big[\phantom{\,}^{\mu} P^{\nu}\big]
       \frac{1}{4\eta\eta'}
   \Big\{\partial^2
       \Big[
            \Theta(\Delta\tau^2)\Theta(\Delta\eta)\ln|\Delta\tau^2|
       \Big]
     + \Big(
            3 \!-\! \ln(4\eta\eta') \!-\! \frac{1}{\tt s}
       \Big)
       \partial^2
       \Big[
            \Theta(\Delta\tau^2)\Theta(\Delta\eta)
       \Big]
  \Big\}
\nonumber\\
    &-& \big[\phantom{\,}^{\mu} P^{\nu}\big]
    \frac{\Theta(\Delta\tau^2) \Theta(\Delta\eta)}
         {(\bar y-4)\eta^2\eta'^2}
\nonumber\\
    &+& \big[\phantom{\,}^{\mu}\bar P^{\nu}\big]
      \frac{\Theta(\Delta\tau^2) \Theta(\Delta\eta)}{\eta^2\eta'^2}
      \Big\{
            \frac{1}{2}\Big[
                            \ln\Big|\frac{\bar y}{4}\Big|
                          + \ln\Big(1-\frac{\bar y}{4}\Big)
                       \Big]
         +  \Big[
                 \frac{\bar y-2}{\bar y-4}+2-\frac{1}{2\tt s}
            \Big]
      \Big\}
\bigg\}
\nonumber\\
    &+& O({\tt s})
\,.
\label{eq - ret vac pol}
\end{eqnarray}

\section{Effective field equation and photon mass}
\label{Effective field equation and photon mass}

 An effective  field equation for photons in de Sitter space-time
can be derived by the variation of the Schwinger-Keldysh effective action.
The result is the following non-local
equation~\cite{ProkopecTornkvistWoodard:2002AoP},
\begin{equation}
     \eta^{\nu\rho}\eta^{\mu\sigma}\partial_\nu F_{\rho\sigma}
   + \int d^4x'[^{\mu}\Pi^{r,\nu}_{ren}](x,x')A_\nu(x')
   + O(A^3)
   = 0
\,.
\label{eq - effective field equation em ipi ret}
\end{equation}
For simplicity here we shall approximate the retarded vacuum polarization
tensor~(\ref{eq - ret vac pol}) by its leading order
$O({\tt s}^{-1})$ contribution,
\begin{equation}
    [^{\mu}\Pi^{r,\nu}_{ren}](x,x')
  \simeq   \frac{\alpha}{8\pi^2\tt s}
      \Big[\big[\phantom{\,}^{\mu} P^{\nu}\big]
           \frac{\partial^2\Theta(\Delta\tau^2)
                           \Theta(\Delta\eta)}
                {\eta\eta'}
    - \big[\phantom{\,}^{\mu} \bar P^{\nu}\big]
        \frac{2\Theta(\Delta\tau^2)\Theta(\Delta\eta)}
             {\eta^2\eta'^2}
       \Big]
\label{Pi:ren-ret}
\,,
\end{equation}
which we shall use to study the effective field
equation~(\ref{eq - effective field equation em ipi ret}).
This is justified provided $|m_\phi^2+\xi R|\ll H^2$.
Upon neglecting the $O(A^3)$ contributions
in (\ref{eq - effective field equation em ipi ret}),
we seek the solutions in the form,
\begin{equation}
    A_{\nu}(x')=\varepsilon_{\nu}(\vec{k},\eta')e^{i\vec{k}\cdot\vec{x}'}
\,,
\label{eq - A ansatz}
\end{equation}
where
\begin{equation}
    \Big(\partial'_0-\frac{2}{\eta'}\Big)\varepsilon_0(\vec{k},\eta')
  = i\vec k\cdot \vec \varepsilon(\vec{k},\eta')
\,,
\label{eq - eps gauge}
\end{equation}
which is equivalent to the generalized Lorentz 
gauge~(\ref{Lorentz gauge:generalisation}). This gauge is obtained by
requiring that divergence of the Proca equation~(\ref{eq - eq of motion A})
in de Sitter space vanishes, as implied by gauge invariance.
Inserting Eqs.~(\ref{Pi:ren-ret}) and ~\ref{eq - A ansatz}) into~(\ref{eq -
effective field equation em ipi ret}) and evaluating the integral
gives the following Proca equation (see Appendix \ref{sec -
integral in field equation} for the derivation)
\begin{equation}
     \eta^{\rho\nu}\partial_{\nu}(\partial_{\rho}A_{\mu}
   - \partial_{\mu}A_{\rho})
   - a^2 m_\gamma^2 A_{\mu}
   = 0
\,,
\label{eq - eq of motion A}
\end{equation}
with the photon mass given by
\begin{eqnarray}
    m_{\gamma}^2 &=& \frac{\alpha H^2}{\pi\tt s}+O({\tt s}^0)
\nonumber\\
                 &=& \frac{3\alpha H^4}{\pi(m_\phi^2+ \xi R)}
                  + O\Big(\Big(\frac{m_\phi^2+ \xi R}{H^2}\Big)^0\Big)
\,,
\label{photon mass}
\end{eqnarray}
where $R$ is the curvature scalar, which in de Sitter space,
$R=12H^2$, and $\xi$ specifies the coupling of the scalar field to
gravity. Thus photons that couple to light minimally coupled
scalar particles acquire a mass in inflation. A remarkable feature
of this result is that, even though $\alpha \ll 1$, such that the
one-loop approximation is justified,  the photon mass may be much
larger than the Hubble parameter.

\subsection{On the speed of light in inflation}
\label{sec - speed of light}

We can use the Proca equation (\ref{eq - eq of motion A}) to study the propagation of light in inflation.
We seek a solution of the form,
\begin{equation}
    A_{\mu}(x)=\varepsilon_{\mu}(\vec{k},\eta)e^{i\vec{k}\cdot\vec{x}}
\,,
\end{equation}
where
$\varepsilon_{\mu}(\vec{k},\eta)=(\varepsilon_{0}(\vec{k},\eta),\vec{\varepsilon}\,(\vec{k},\eta))$.
Thus the $\mu=0$ component of (\ref{eq - eq of motion A}) can be written as
\begin{equation}
    \varepsilon_0(\vec{k},\eta)=-\frac{i\partial_0\vec{k}\cdot\vec{\varepsilon}\,(\vec{k},\eta)}
    {\vec{k}^2+m^2_\gamma a^2},
\end{equation}
which tells us that the zeroth component of the photon field traces
the spatial components. 
Using the gauge~(\ref{eq - eps gauge}) and decomposing
$\vec{\varepsilon}\;$ into the longitudinal and transverse parts
\begin{equation}
    \vec{\varepsilon}_L\equiv\frac{(\vec{k}\cdot\vec{\varepsilon})}{\vec{k}^2}\vec{k}\mspace{27mu}
    ,\mspace{27mu}\vec{\varepsilon}_T\equiv\vec{\varepsilon}-\vec{\varepsilon}_L,
\end{equation}
the spatial components of the Proca equation~(\ref{eq - eq of motion A}) can
be recast as
\begin{eqnarray}
  (\partial_0^2+\vec{k}^2+m_{\gamma}^2a^2)\,\vec{\varepsilon}_T &=& 0
\label{eq - eom epsT}\\
  \Big(\partial_0^2+\vec{k}^2+m_{\gamma}^2a^2
    + \frac{2Ha\vec{k}^2}{\vec{k}^2+m^2_\gamma
    a^2}\partial_0\Big)\,\vec{\varepsilon}_L &=& 0
\,.
\label{eq - eom epsL}
\end{eqnarray}
Consider now a longitudinally polarised photon
$\vec A_L(x) = \vec{\varepsilon}_L e^{i\vec k\cdot \vec x}$, 
$\vec{\varepsilon}_L = \vec{\varepsilon}_L(\vec k,\eta)$.
While the magnetic field is trivially equal to zero, 
$\vec B_L = a^{-2}\nabla\times \vec A_L = 0$, the electric field does not
vanish,
\begin{equation}
  \vec E_L (x) = - \frac{1}{a^2}
                 \,\frac{1}{1+(k/am_\gamma)^2} \,
                 \partial_\eta \vec{\varepsilon}_L\; e^{i\vec k\cdot \vec x}
\,. 
\label{electric-field:longitudinal}
\end{equation}
 From this result one can nicely see that, in the limit when 
$m_\gamma \rightarrow 0$, the electric field vanishes as
$\vec E_L \propto m_\gamma^2$, rendering the longitudinal polarization
unphysical. On the other hand, when $m_\gamma$ grows and becomes comparable to,
or larger than, $k/a$, the amplitude of $E_L$ is unsuppressed.

Writing $\vec{\varepsilon}_T$, $\vec{\varepsilon}_L$ as
\begin{equation}
    \vec{\varepsilon}_{(T,L)}(\vec{k},\eta)=\vec{c}_{(T,L)}(\vec{k})\,\alpha_{(T,L)}(\vec{k},\eta)\,\exp\Big(
    - i\int_{\eta_0}^\eta\omega_{(T,L)}(\vec{k},\eta')d\eta'\Big),
    \label{eq - ansatz epsT,L}
\end{equation}
where $\alpha_{(T,L)}$ and $\omega_{(T,L)}$ are real by
construction, one obtains from (\ref{eq - eom epsT}) and (\ref{eq
- eom epsL})
\begin{equation}
    \frac{\omega'_T}{\omega_T}+2\frac{\alpha'_T}{\alpha_T}=0\mspace{27mu},
    \mspace{27mu}\frac{\alpha''_T}{\alpha_T}+\vec{k}^2+m_\gamma^2
    a^2=\omega_T^2,
    \label{eq - transverse eom}
\end{equation}
and
\begin{align}
    &\frac{\omega'_L}{\omega_L}+2\frac{\alpha'_L}{\alpha_L}=-\frac{2Ha\vec{k}^2}{\vec{k}^2+m_{\gamma}^2a^2},
    \label{eq - longitudinal eom 1}\\
    \frac{\alpha''_L}{\alpha_L}+&\vec{k}^2+m_\gamma^2
    a^2+\frac{2Ha\vec{k}^2}{\vec{k}^2+m_{\gamma}^2a^2}\frac{\alpha'_L}{\alpha_L}=\omega_L^2,
    \label{eq - longitudinal eom 2}
\end{align}
where the {\it prime} denotes a derivative with respect to $\eta$
($\,^\prime\equiv(d/d\eta)$).
Solving the first equation of (\ref{eq
- transverse eom}) and Eq.~(\ref{eq - longitudinal eom 1}) gives
\begin{equation}
    \alpha_T=\frac{1}{\sqrt{\omega_T}}\mspace{27mu},\mspace{27mu}\alpha_L=\frac{1}{\sqrt{\omega_L}}
    \exp\Big(-\int_{\eta_0}^\eta\frac{Ha'\vec{k}^2}{\vec{k}^2+m_{\gamma}^2a'^2}d\eta'\Big).
    \label{eq - solutions alphaT,L}
\end{equation}
Performing the integration in the second equality gives
\begin{equation}
\alpha_L=\frac{1}{\sqrt{\omega_L}}\sqrt{\frac{m_\gamma^2+(\frac{\vec k}{a})^2}{m_\gamma^2+(\frac{\vec k}{a_0})^2}},
\end{equation}
where $a_0\equiv a(\eta_0)$. Thus in the oscillatory regime, where
one can find real solutions for $\omega_L$, longitudinal modes
become suppressed compared to transverse modes during inflation,
unless they are non-relativistic, which implies $\|\vec k\|/a_0
\ll m_\gamma$. Upon plugging (\ref{eq - solutions alphaT,L}) into
the second equation of (\ref{eq - transverse eom}) and
Eq.~(\ref{eq - longitudinal eom 2}), we find the following equations
for $\omega_T=\omega_T(\vec k,\eta)$ and $\omega_L=\omega_L(\vec
k,\eta)$
\begin{align}
   \omega_T^2 &= \vec k^2 + a^2 m_\gamma^2
            + \frac 34\Big(\frac{\omega_T^\prime}{\omega_T}\Big)^{2}
            - \frac 12\frac{\omega_T^{\prime\prime}}{\omega_T}\,,
    \label{omegaT:eom}\\
    \omega_L^2 &= \vec k^2 + a^2 m_\gamma^2
            + \frac 34\Big(\frac{\omega_L^\prime}{\omega_L}\Big)^{2}
            - \frac
            12\frac{\omega_L^{\prime\prime}}{\omega_L}+
            \frac{H^2a^2\vec k^2(m_\gamma^2a^2-2\vec k^2)}{(\vec{k}^2+m_{\gamma}^2a^2)^2}\,.
    \label{omegaL:eom}
\end{align}
In adiabatic limit, when
\begin{equation}
\omega_{(T,L)}^2 \gg \omega_{(T,L)}^\prime \,,\qquad
\omega_{(T,L)}^3 \gg \omega_{(T,L)}^{\prime\prime} \,,
\label{adiabatic-conditions}
\end{equation}
Eqs.~(\ref{omegaT:eom}) and (\ref{omegaL:eom}) can be iteratively
solved. The conditions~(\ref{adiabatic-conditions}) are satisfied for
non-relativistic photons ($k_{\rm ph} \equiv \|\vec k\|/a \ll m_\gamma$),
when
\begin{equation}
\qquad\qquad\quad
      H  \ll m_\gamma
\qquad\qquad\qquad
    ({\tt non\!\!-\!\!relativistic\;\; limit}:\; k_{\rm ph} \ll m_\gamma)
\,.
\label{ac:NR-limit}
\end{equation}
Then the last term of Eq.~(\ref{omegaL:eom}) can also be
neglected. In the relativistic case ($k_{\rm ph} \gg m_\gamma$) we
get from (\ref{adiabatic-conditions}) $(k_{\rm ph} \gg
(Hm_\gamma)^\frac 12, \; (Hm_\gamma^2)^\frac 13)$, but this is not
sufficient for longitudinal modes. We need the stronger condition
\begin{equation}
 \begin{array}{cc}
   k_{\rm ph} \gg  (Hm_\gamma)^\frac 12 & {\rm (transverse\;\; modes)}    \cr
   k_{\rm ph} \gg  H  \quad             & {\rm (longitudinal\;\; modes)}
  \end{array}
\Biggr\}
\qquad ({\tt relativistic\;\; limit}:\; k_{\rm ph} \gg m_\gamma)
\,,
\label{ac:R-limit}
\end{equation}
for the adiabatic approximation to work, and the last term in
Eq.~(\ref{omegaL:eom}) can be neglected.

When the conditions~(\ref{ac:NR-limit}) or (\ref{ac:R-limit})
are met, one finds from Eqs.~(\ref{omegaT:eom}--\ref{omegaL:eom})
to leading order in derivatives
\begin{equation}
   \omega_{(T,L)} \simeq \omega_{0}
\,,\qquad
  \omega_{0}  = \sqrt{\vec k^2 + a^2 m_\gamma^2}
\,.
\label{omega:adiabatic}
\end{equation}
The propagation speed of massive photons is then given by the group velocity
\begin{equation}
   \vec v_{\rm group} = \frac{d\omega_{\rm ph}}{d\vec{k}_{\rm ph}}
                 \simeq \frac{\vec{k}_{\rm ph}}{\omega_{\rm ph}}
\,,
\end{equation}
which is always smaller than the speed of light in {\it vacuo}.
When $\|\vec k_{\rm ph}\| \ll m_\gamma$, $v_{\rm group}$ can be $\ll 1$.
Here we defined the physical frequency,
$\omega_{\rm ph} \simeq \omega_{0}/a$, and the physical wave
vector, $\vec{k}_{\rm ph} = \vec k/a $.

When adiabatic approximation~(\ref{adiabatic-conditions})
breaks down, one has to solve for
$\omega_{(T,L)} = \omega_{(T,L)}(\vec k,\eta)$
exactly. As an example, adiabatic approximation for longitudinal
relativistic modes breaks down when $k_{\rm ph}$ approaches $2H$, even
if $k_{\rm ph} \gg (Hm_\gamma)^\frac 12$, is satisfied, which means that
the derivative terms in Eqs.~(\ref{omegaT:eom}--\ref{omegaL:eom})
can be neglected.
In this case, and when $m_\gamma\ll k_{\rm ph}, H$,
Eqs.~(\ref{omegaL:eom}) and~(\ref{eq - eom epsL})
take the following form
\begin{align}
    &\omega_L^2 \simeq \vec k^2 - 2H^2a^2,\label{eq - omegaL m to 0 limit}\\
    (\partial_0^2&+\vec{k}^2-2a^2H^2)a\vec{\varepsilon}_L=0.\label{eq - eom epsL m to 0 limit}
\end{align}
 From~(\ref{eq - omegaL m to 0 limit}) one would not expect
oscillatory behavior for $k_{\rm ph}<2H$, but solving
(\ref{eq - eom epsL m to 0 limit}) gives
\begin{equation}
    \vec{\varepsilon}_L(\vec k,\eta)= \hat{\vec k}
                                       \sqrt{k^{-2}+\eta^2}
\; {\rm exp}\Big\{-i\Big(
                        k\eta+\arctan\big(\frac{1}{k\eta}\big)
                  \Big)
          \Big\},
\end{equation}
which is oscillatory. Here we used $k\equiv \|\vec k\|$ and  
$\hat{\vec k} = \vec k/k$.
Comparing this expression to
Eq.~(\ref{eq - ansatz epsT,L}) one obtains
$\omega_L=k-k/({1+\frac{k^2}{H^2a^2}})$
for the dispersion relation, such that the group velocity becomes
\begin{equation}
 v_{\rm group} = 1 - \frac{1-({k_{\rm ph}}/{H})^2}
                          {\big[1+(k_{\rm ph}/H)^2\big]^2}
\,.
\end{equation}
This implies that subhorizon photons ($k_{\rm ph} > H$) propagate
superluminally, while superhorizon photons ($k_{\rm ph} < H$)
are subluminal. When one considers propagation of relativistic photons 
on subhorizon scales, one finds that longitudinal (transverse) 
photons propagate superluminally (subluminally).
This phenomenon is similar to 
birefringence~\cite{DrummondHathrell:1979,Prokopec:2001},
where one of the tranverse polarizations may propagate superluminally, 
and another subluminally.

\section{Physical implications of our results}
\label{Physical implications of our results}

In this section we study the cosmological consequences of a
massive photon in inflation.
In the Introduction we argued that a large photon mass during
inflation~(\ref{photon mass}) may dramatically influence
the photon dynamics, perhaps the most striking being
the speed of its propagation: the scalar vacuum fluctuations
act as an `\ae ther', which drags photons, and consequently may dramatically
slow down propagation of light.

 Another interesting consequence may be creation of magnetic fields
on cosmological scales with potentially observable
magnitudes~\cite{DavisDimopoulosProkopecTornkvist:2000,
DimopoulosProkopecTornkvistDavis:2001,ProkopecWoodard:2003ajp}.
By following the derivation of Ref.~\cite{ProkopecWoodard:2003ajp},
we arrive at the following estimate for the (volume-averaged) magnetic
field on a (comoving) scale $\ell_c$~\footnote{This estimate is by a factor
$(8\pi)^{1/2}$ times larger than the result in~\cite{ProkopecWoodard:2003ajp}.
The difference can be traced to the factor $(8\pi)^{-1}$ in the definition
of the electromagnetic energy density in the Gaussian system of units.}:
\begin{align}
    B(t,\ell_c) & = \Big(
                      \frac{3\alpha}{\pi z_{\rm eq}}
                 \Big)^{\frac{1}{4}}
                 \Big(
                      \frac{H_0}{2\pi m_\phi c^3}
                 \Big)^{\frac{1}{2}}
                      \frac{H\hbar(1+z)^2}{\ell_c}
\mspace{27mu}(\text{in the Gaussian system}),
\label{eq - seed field gauss}\\
    B(t,\ell_c) & = \sqrt{\frac{\mu_0}{4\pi}}\Big(
                      \frac{3\alpha}{\pi z_{\rm eq}}
                 \Big)^{\frac{1}{4}}
                 \Big(
                      \frac{H_0}{2\pi m_\phi c^3}
                 \Big)^{\frac{1}{2}}
                       \frac{H\hbar(1+z)^2}{\ell_c}
\mspace{27mu}(\text{in SI})
\label{eq - seed field SI}
\,,
\end{align}
where we have reinserted the physical constants. Here $z_{\rm
eq}\approx3200$ denotes the redshift at the matter-radiation
equality, $H_0 \simeq 2.3 \times 10^{-18}~{\rm Hz}$ is the present
time Hubble parameter (which corresponds to $H_0 \simeq
71$\,km/s/Mpc), $H \approx 10^{13}~{\rm GeV}/\hbar$ is the Hubble
parameter during inflation, the fine structure constant 
$\alpha = 1/137$, and $\mu_0$ is the
magnetic permeability of vacuum. The result~(\ref{eq - seed field
gauss}) is derived by making the assumption that $1/2$ of the
energy stored in vacuum fluctuations during inflation is converted
into the magnetic energy at the second horizon crossing during
radiation era, when also the photon mass is assumed to vanish
nonadiabatically. 
For an alternative derivation we refer to
Refs.~\cite{DavisDimopoulosProkopecTornkvist:2000,
DimopoulosProkopecTornkvistDavis:2001}, where continuous matching
of the field amplitude and its derivative at the
inflation-radiation transition are employed, and the photon is
assumed to become massless at the inflation-radiation transition.

Provided the scalar field is sufficiently light,
naively it seems that the field 
strength~(\ref{eq - seed field gauss}-\ref{eq - seed field SI}) can be 
significantly larger than the one obtained in 
Ref.~\cite{ProkopecWoodard:2003ajp}, where the photon dynamics coupled to 
a massless minimally coupled scalar was considered.
Taking account of the more recent results obtained
in~\cite{ProkopecWoodard:2003photons} however, 
which correctly treats the late time asymptotic dynamics of the photon field
in inflation, we conclude that one would get equally strong magnetic fields
from photons coupled to a minimally coupled massless scalar,
provided inflation lasts a sufficiently long time.

 Evaluating Eq.~(\ref{eq - seed field gauss}) with $m_\phi \simeq
100~{\rm GeV}$ \footnote{The scalar mass of electroweak scale
implies a very large (super-Planckian) photon mass, $m_\gamma \sim
10^{22}~{\rm GeV}$. A more realistic estimate for the photon mass
in this case would be $m_\gamma \sim M_{\rm Pl} \sim 10^{19}~{\rm
GeV}$, which formally corresponds to $m_\phi \sim 300~{\rm TeV}$.
When the Planck mass is used in~(\ref{eq - seed field gauss}), one
gets about two orders of magnitude weaker seed field.}, we get
\begin{equation}
     B(t,\ell_c)\approx 10^{-28}\frac{(1+z)^2}{\ell_c/10kpc}\text{Gauss}.
\end{equation}
Assuming galaxy formation at $z\sim 10$ and an amplification by a
factor $\sim80$ through field compression during the collapse of
the proto-galaxy~\cite{GrassoRubinstein:2000},
one gets field strengths of approximately
%
   $
    B \sim 10^{-24}\text{ G},
   $
%
for galactic magnetic fields after structure formation at the
scales relevant for galactic dynamos. $\ell_c\sim 10\text{ kpc}$ has
been used, which corresponds roughly to $\sim100\text{ pc}$
physical length after the field compression at $z\sim 10$. This
is most likely sufficient to seed a dynamo mechanism, which generates the
micro-Gauss galactic magnetic fields observed
today~\cite{ZeldovichRuzmaikinSokoloff:1983,KulsrudCowleyGruzinovSudan:1997}.

\acknowledgements

 We thank Richard Woodard for very useful discussions and suggestions,
which were instrumental for bringing the paper to its current form.
We would like to thank Michael G. Schmidt for engaging guidance of 
E.P.'s diploma thesis~\cite{Puchwein:2003}, 
a product of which is the work presented here. 

\appendix

\section{Expanding the scalar propagator
}
\label{sec - app prop expansion}

We want to expand the scalar Feynman propagator
\begin{equation}
    i\Delta(x,x')=\frac{\Gamma(\frac{D-1}{2}+\nu)
    \Gamma(\frac{D-1}{2}-\nu)}{(4\pi)^{\frac{D}{2}}
    \Gamma(\frac{D}{2})}H^{D-2}\;
{}_2F_1\Big(\frac{D-1}{2}+\nu,\frac{D-1}{2}-\nu,\frac{D}{2},1-\frac{y}{4}\Big),
\end{equation}
in the modified de Sitter length function $y$, {\tt s} and
$\varepsilon\equiv 4-D$ around $y=0$, ${\tt s}=0$ and $\varepsilon=0$, where
\begin{equation}
    \nu\equiv\frac{D-1}{2}- {\tt s}
        = \frac{D-1}{2}-\frac{1}{D-1}\frac{m_\phi^2+ \xi R}
    {H^2}+O\left(\Big(\frac{m_\phi^2+ \xi R}{H^2}\Big)^2\right).
    \label{eq - nu app}
\end{equation}
Using the following transformation formula
\begin{align}
    {}_2F_1&
\left(\frac{D-1}{2}+\nu,\frac{D-1}{2}-\nu,\frac{D}{2},1-\frac{y}{4}\right)
\nonumber \\
    &=\frac{\Gamma(\frac{D}{2})\Gamma(1-\frac{D}{2})}
    {\Gamma(\frac{1}{2}-\nu)\Gamma(\frac{1}{2}+\nu)}\;
{}_2F_1\left(\frac{D-1}{2}+\nu,\frac{D-1}{2}-\nu,\frac{D}{2},\frac{y}{4}
       \right)
\nonumber\\
    &+\frac{1}{(\frac{y}{4})^{\frac{D}{2}-1}}\frac{\Gamma(\frac{D}{2})\Gamma(\frac{D}{2}-1)}
    {\Gamma(\frac{D-1}{2}-\nu)\Gamma(\frac{D-1}{2}+\nu)}\;
{}_2F_1\left(\frac{1}{2}+\nu,\frac{1}{2}-\nu,2-\frac{D}{2},\frac{y}{4}
       \right)
,
\end{align}
and the series expansion of the hypergeometric function
\begin{equation}
{}_2F_1(a,b,c,z)
  = \frac{\Gamma(c)}{\Gamma(a)\Gamma(b)}
    \sum_{n=0}^{\infty}\frac{\Gamma(a+n)\Gamma(b+n)}{\Gamma(c+n)}\frac{z^n}{n!}
,
\end{equation}
one can easily derive
\begin{align}
    i\Delta(x,x')&=\frac{H^{D-2}}{(4\pi)^{\frac{D}{2}}}
 \bigg[\frac{\Gamma(1-\frac{D}{2})\Gamma(\frac{D}{2})}
    {\Gamma(\frac{1}{2}-\nu)\Gamma(\frac{1}{2}+\nu)}\sum_{n=0}^{\infty}
    \frac{\Gamma(\frac{D-1}{2}+\nu+n)\Gamma(\frac{D-1}{2}-\nu+n)}{\Gamma(\frac{D}{2}+n)}
    \frac{(\frac{y}{4})^n}{n!} \nonumber \\
    +&\frac{1}{(\frac{y}{4})^{\frac{D}{2}-1}}\frac{\Gamma(\frac{D}{2}-1)\Gamma(2-\frac{D}{2})}
    {\Gamma(\frac{1}{2}-\nu)\Gamma(\frac{1}{2}+\nu)}\sum_{n=0}^{\infty}
    \frac{\Gamma(\frac{1}{2}+\nu+n)\Gamma(\frac{1}{2}-\nu+n)}{\Gamma(2-\frac{D}{2}+n)}
    \frac{(\frac{y}{4})^n}{n!}
 \bigg].
\end{align}
Making use of
$\Gamma(\frac{1}{2}+b)\Gamma(\frac{1}{2}-b)=\frac{\pi}{\cos(\pi
b)}$, $\Gamma(b)\Gamma(1-b)=\frac{\pi}{\sin(\pi b)}$ and inserting
in the first equality in (\ref{eq - nu app}) one obtains
\begin{align}
    i\Delta(x,x')
  = \frac{H^{D-2}}{(4\pi)^{\frac{D}{2}}}
    & \frac{\cos(\pi(\frac{D-1}{2}- {\tt s}))}{\sin(\pi\frac{D}{2})}
 \bigg[\sum_{n=0}^{\infty}
        \frac{\Gamma(D-1- {\tt s} + n)\Gamma({\tt s}+n)}{\Gamma(\frac{D}{2}+n)}
    \frac{(\frac{y}{4})^n}{n!} \nonumber \\
    -&\frac{1}{(\frac{y}{4})^{\frac{D}{2}-1}}\sum_{n=0}^{\infty}
    \frac{\Gamma(\frac{D}{2}- {\tt s} + n)\Gamma(1-\frac{D}{2}+ {\tt s} +n)}
         {\Gamma(2-\frac{D}{2}+n)}
    \frac{(\frac{y}{4})^n}{n!}
 \bigg].
\end{align}
This is still formally exact in  {\tt s} and $\varepsilon$. Upon pulling
the $n=0$ term out of the first sum, the $n=0$ and $n=1$ terms out
of the second sum and shifting the second sum we find,
\begin{align}
  i\Delta(x,x')=\frac{H^{D-2}}{(4\pi)^{\frac{D}{2}}}
               & \frac{\cos(\pi(\frac{D-1}{2}- {\tt s}))}{\sin(\pi\frac{D}{2})}
 \bigg[\sum_{n=1}^{\infty}\frac{\Gamma(D-1-{\tt s}+n)\Gamma({\tt s}+n)}
                                 {\Gamma(\frac{D}{2}+n)}
    \frac{(\frac{y}{4})^n}{n!} \nonumber \\
    -&\big(\frac{y}{4}\big)^{2-\frac{D}{2}}\sum_{n=1}^{\infty}
    \frac{\Gamma(\frac{D}{2}-{\tt s}+n+1)\Gamma(2-\frac{D}{2}+{\tt s}+n)}
         {\Gamma(3-\frac{D}{2}+n)}
    \frac{(\frac{y}{4})^n}{(n+1)!} \nonumber \\
    +&\frac{\Gamma(D-1-{\tt s})\Gamma({\tt s})}{\Gamma(\frac{D}{2})}
    -\frac{1}{(\frac{y}{4})^{\frac{D}{2}-1}}
     \frac{\Gamma(\frac{D}{2}-{\tt s})\Gamma(1-\frac{D}{2}+{\tt s})}
          {\Gamma(2-\frac{D}{2})}
    \nonumber \\
    -&\big(\frac{y}{4}\big)^{2-\frac{D}{2}}
      \frac{\Gamma(\frac{D}{2}-{\tt s}+1)\Gamma(2-\frac{D}{2}+{\tt s})}
           {\Gamma(3-\frac{D}{2})}
\bigg]
\,.
\end{align}
Then expanding in {\tt s} gives
\begin{align}
    i\Delta(x,x')=\frac{H^{2-\varepsilon}}{(4\pi)^{2-\frac{\varepsilon}{2}}}
&\bigg[
    \frac{\Gamma(3-\varepsilon)}{\Gamma(2-\frac{\varepsilon}{2})}
    \Big\{\frac{1}{{\tt s}}
       +  \pi\cot\big(\frac{\pi\varepsilon}{2}\big)
                  -\gamma_{\rm E}
    -\psi(3-\varepsilon)+C({\tt s},\varepsilon)
    \Big\}
\nonumber\\
    -&\big(\frac{y}{4}\big)^{\frac{\varepsilon}{2}}
        \frac{\Gamma(3-\frac{\varepsilon}{2})}{\frac{\varepsilon}{2}}
  \Big\{1+{\tt s}\Big(\pi\cot\big(\frac{\pi\varepsilon}{2}\big)
                   -  \psi\big(3-\frac{\varepsilon}{2}\big)
                   +  \psi(\frac{\varepsilon}{2})
                 \Big)
  \Big\}
     + \frac{1}{(\frac{y}{4})^{1-\frac{\varepsilon}{2}}}
           \Gamma\big(1-\frac{\varepsilon}{2}\big)
\nonumber\\
    +&\sum_{n=1}^{\infty}
   \Big\{{\tt s}
        \frac{2-n(n+2)\ln(\frac{y}{4})}{n^2}
    + O(\varepsilon)
   \Big\}\big(\frac{y}{4}\big)^{n}
\bigg]
  + O({\tt s}^2)
\,,
\label{eq - app expanded prop}
\end{align}
where $\gamma_{\rm E}\approx0.577$ is the Euler-Mascheroni constant, $\psi$
is defined by $\psi(z)\equiv \Gamma'(z)/\Gamma(z)$ and
$C({\tt s},\varepsilon)$ is an order {\tt s} term that is independent on the
coordinates. The $O(\varepsilon)$ term in the infinite sum, will
not be needed for the regularization of the vacuum polarization.

\section{Feynman rules of scalar QED in position space}

\label{sec - feynman rules sQED}

The position space Feynman rules of scalar QED in an arbitrary
D-dimensional space with metric $g^{\mu\nu}$ are\\\\
\raisebox{-0.32in}{\includegraphics[scale=0.3]{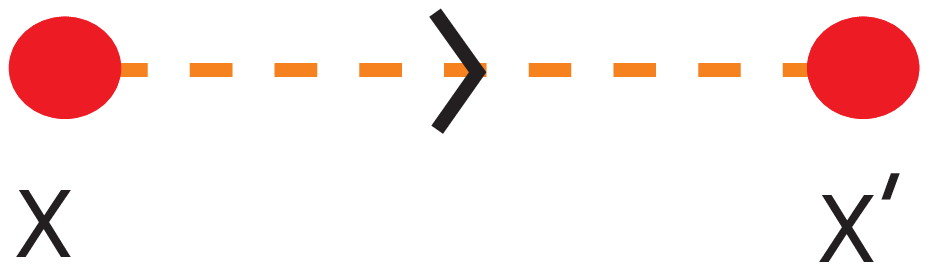}}
$\quad=i\Delta(x,x')$,\\
\raisebox{-0.6in}{\includegraphics[scale=0.3]{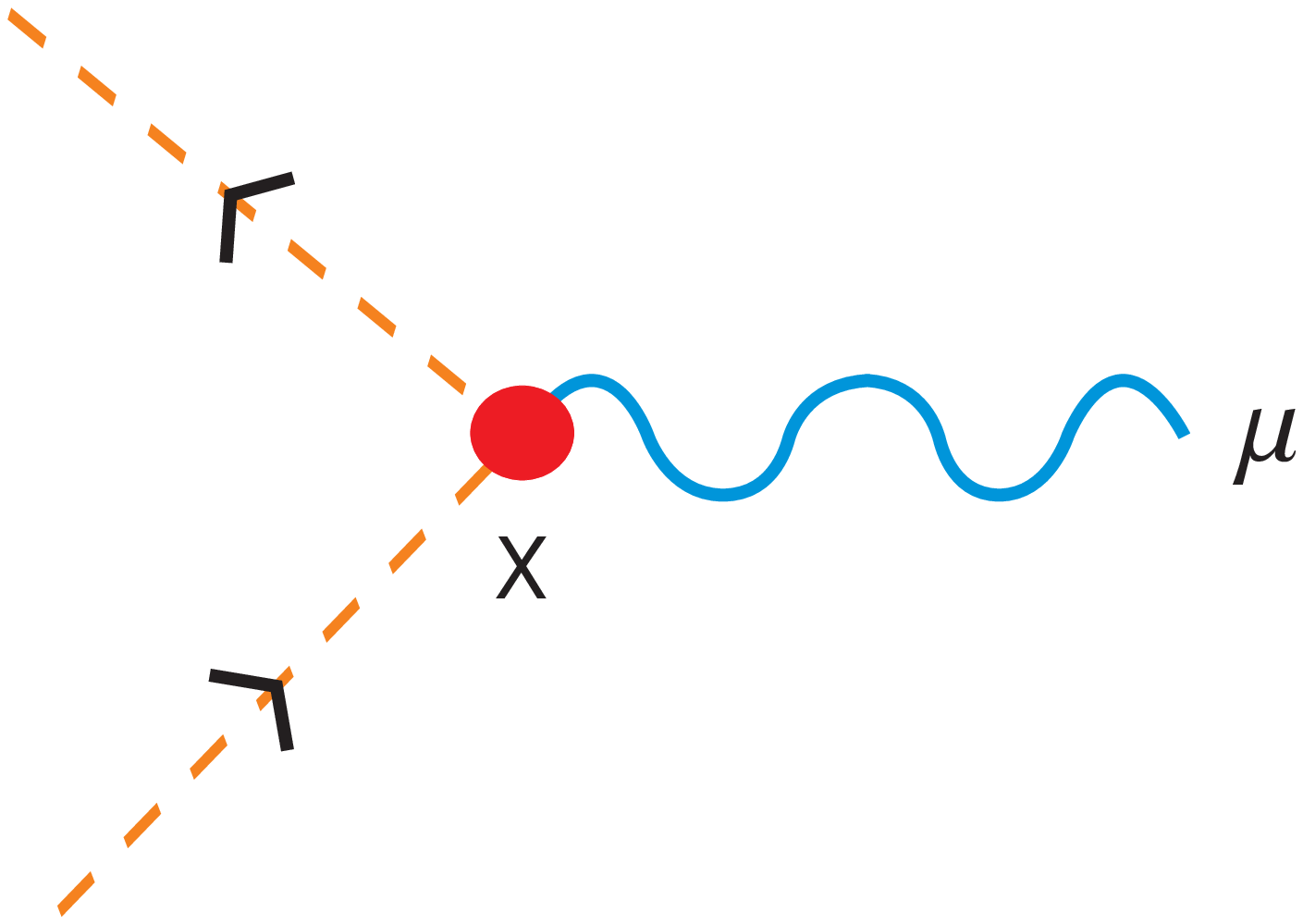}}
$\quad=e\sqrt{-g(x)}g^{\mu\sigma}(x)(\partial^{\,\rm out}_{\sigma}
       -\partial^{\,\rm in}_{\sigma})$,\\\\
\raisebox{-0.32in}{\includegraphics[scale=0.3]{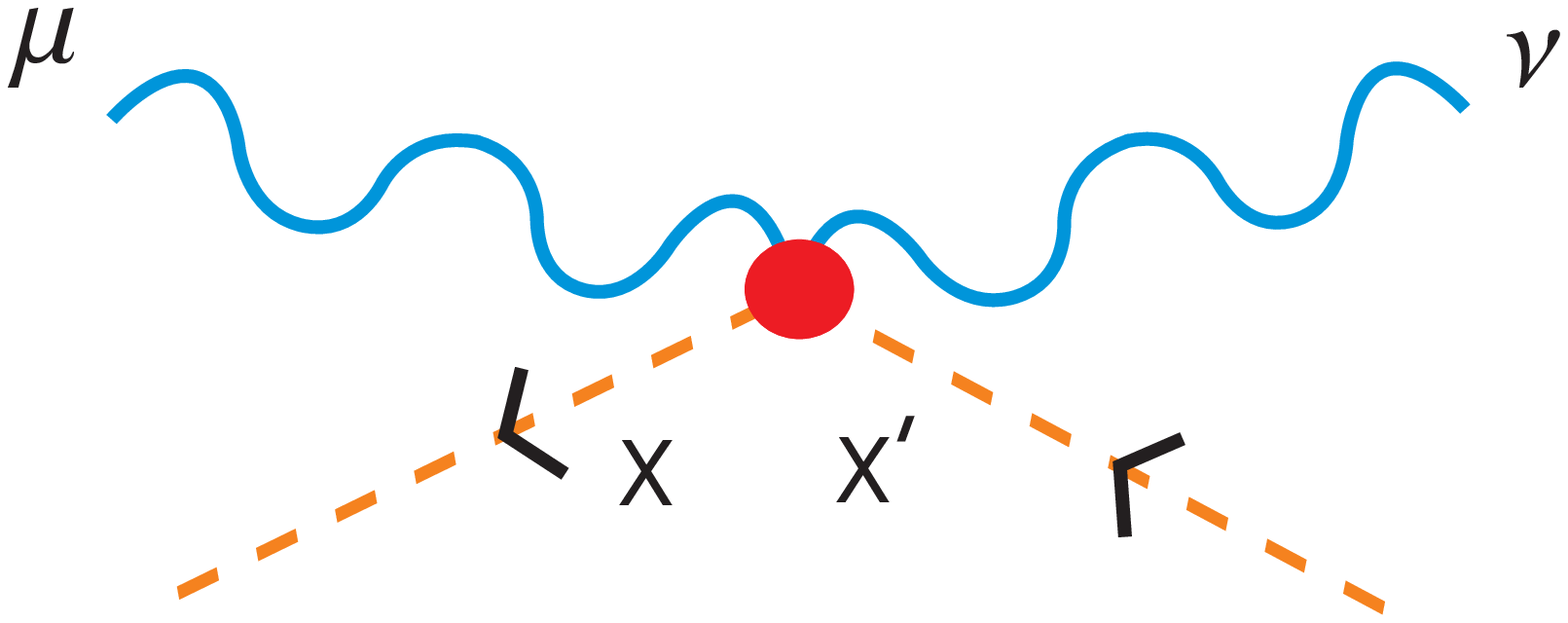}}
$\quad=-2ie^2\sqrt{-g(x)}g^{\mu\nu}(x)\delta^D(x-x')$,\\\\\\
\raisebox{-0.32in}{\includegraphics[scale=0.3]{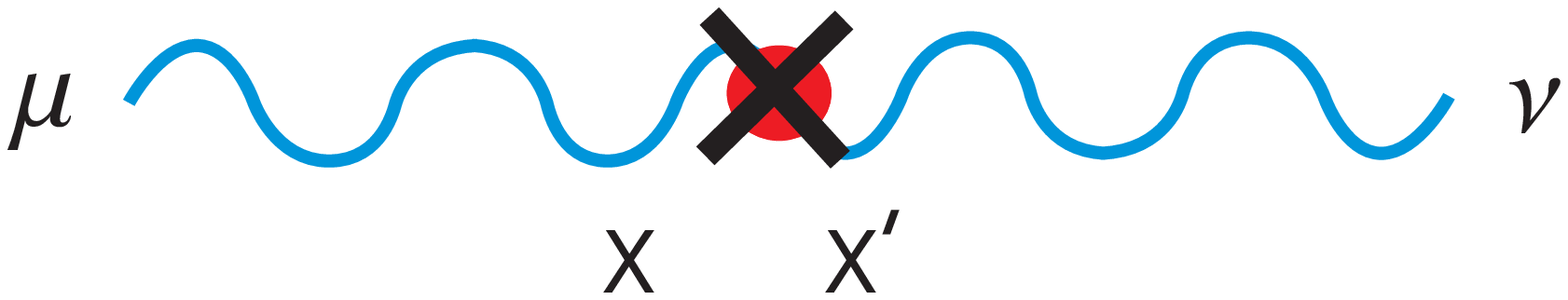}}
$=-i\delta Z\partial_{\rho}(\sqrt{-g(x)}
 [g^{\mu\nu}(x)g^{\rho\sigma}(x)-g^{\mu\sigma}(x)g^{\nu\rho}(x)]
 \partial_{\sigma}'\delta^D(x-x'))$,\\
where $g\equiv\det[g_{\mu\nu}]$, $\partial^{\,\rm in}_{\sigma}$ acts only
on the incoming and $\partial^{\,\rm out}_{\sigma}$ only on the outgoing
propagator.  These rules can be derived from the Lagrangean
density (\ref{PhiQED}). In de Sitter D-space $i\Delta(x,x')$ and
$g_{\mu\nu}(x)$ are given by (\ref{eq - prop minimally}) and
(\ref{deSitter:metric}), respectively, and $\sqrt{-g(x)}=a^D$.

\section{The 3-point and 4-point interaction contribution to
         the one-loop vacuum polarization}
\label{sec - vac pol}

 The 3-point and 4-point (seagull)
interaction contributions to the one-loop vacuum polarization 
shown in figure~\ref{fig - vac pol graphs}.(1)
and~\ref{fig - vac pol graphs}.(2) are
\begin{equation}
 i[^{\mu}\Pi^{\nu}]^{(1)}(x,x')
   = - 2ie^2a^{D-2}\eta^{\mu\nu}i\Delta(x,x)\delta^D(x-x'),
\label{eq - app fig.1}
\end{equation}
and
\begin{align}
i[^{\mu}\Pi^{\nu}]^{(2)}(x,x')
   = 2e^2a^{D-2}a'^{D-2}\eta^{\mu\rho}\eta^{\nu\sigma}
    \big[(\partial_{\rho}i\Delta(x,x'))(\partial'_{\sigma}i\Delta(x,x'))-i\Delta(x,x')\partial_{\rho}\partial'_{\sigma}i\Delta(x,x')\big].
    \label{eq - app fig.2}
\end{align}
For $i\Delta(x,x)$ in~(\ref{eq - app fig.1}), Eq.~(\ref{eq - lim prop}) 
will be used and $i\Delta(x,x')$ is given by (\ref{eq -
prop minimally}). The $O(\varepsilon)$ terms, which are not needed for
dimensional regularization, can be dropped because they will not
contribute when taking D to 4 later. Dropping these and all of
$O({\tt s}^2)$ terms, we find for the derivatives of the propagator
\begin{align}
    \partial_{\rho}i\Delta(x,x')=&\big\{\beta f'(y)
  + {\tt s} \frac{H^{2}}{(4\pi)^{2}}
    [g'(y)+h'(y)]\big\}\partial_{\rho}y,\\
    \partial'_{\sigma}i\Delta(x,x')=&\big\{\beta
    f'(y)+{\tt s}\frac{H^{2}}{(4\pi)^{2}}
    [g'(y)+h'(y)]\big\}\partial'_{\sigma}y,\\
    \partial_{\rho}\partial'_{\sigma}i\Delta(x,x')=&\big\{\beta f''(y)
   + {\tt s}\frac{H^{2}}{(4\pi)^{2}}
    [g''(y)+h''(y)]\big\}(\partial_{\rho}y)(\partial'_{\sigma}y)
\nonumber\\
    &+\big\{\beta f'(y)
     +{\tt s}\frac{H^{2}}{(4\pi)^{2}}[g'(y)+h'(y)]\big\}
                 \partial_{\rho}\partial'_{\sigma}y.
\end{align}
$\beta,f,g$ and $h$ are given in Eqs.~(\ref{eq - def beta}--\ref{eq - def h})
in section~\ref{Scalar propagator}.
Note that the order ${1}/{\tt s}$ of $f(y)$ is constant in $y$.
Keeping only terms up to order ${\tt s}^0$, we can write the portion of
(\ref{eq - app fig.2}) within the bracket as
\begin{align}
    \big[(&\partial_{\rho}i\Delta(x,x'))(\partial'_{\sigma}i\Delta(x,x'))
      -  i\Delta(x,x')\partial_{\rho}\partial'_{\sigma}i\Delta(x,x')\big]
\nonumber\\
    &=\big[
            \beta^2(f'^2-ff'')-2\frac{H^{4}}{(4\pi)^{4}}(g''+h'')
      \big] (\partial_{\rho}y)(\partial^{\,\prime}_{\sigma}y)
    +\big[-\beta^2ff'-2\frac{H^{4}}{(4\pi)^{4}}(g'+h')\big]
     \partial_{\rho}\partial'_{\sigma}y.
    \label{eq - app fig.2 bracket}
\end{align}
Differentiating the modified de Sitter length function $y$ gives,
\begin{align}
    \partial_{\rho}y=&aH[y\delta^0_{\rho}+2a'H\Delta x_{\rho}
    +2ia'H \operatorname{sign}(\eta-\eta')\delta^0_{\rho}\delta]
,
\label{eq - derivatives y start}\\
    \partial'_{\sigma}y=&a'H[y\delta^0_{\sigma}-2aH\Delta x_{\sigma}
    -2iaH \operatorname{sign}(\eta-\eta')\delta^0_{\sigma}\delta],\\
    \partial_{\rho}\partial'_{\sigma}y=&aa'H^2[y\delta^0_{\rho}\delta^0_{\sigma}
    -2aH\Delta x_{\sigma}\delta^0_{\rho}
    +2a'H\Delta x_{\rho}\delta^0_{\sigma}
    -2\eta_{\rho\sigma}\nonumber\\
    &-2iaa'H^2|\eta-\eta'|\delta^0_{\rho}\delta^0_{\sigma}\delta
    -4i\delta(\eta-\eta')\delta^0_{\rho}\delta^0_{\sigma}\delta]
\label{eq - derivatives y stop}
\,,
\end{align}
where $\Delta x_\rho \equiv x_\rho - x_\rho^\prime$.
When taking $\delta$ to zero only the final order $\delta$ term
contributes, and only when it multiplies $f'$
\begin{equation}
    -4iaa'H^2\delta(\eta-\eta')\lim_{\delta\rightarrow0}f'(y)\delta
=
    \frac{i}{\beta a^{D-2}}\delta^{D}(x-x')
.
\end{equation}
This contribution subtracts off the temporal component of (\ref{eq
- app fig.1}). Dropping $O({\tt s})$ terms and inserting~(\ref{eq - lim prop})
into~(\ref{eq - app fig.1}),
we find for the remaining part $i[^{\mu}\Pi^{\nu}]^{(1)}$
\begin{equation}
i[^{\mu}\Pi^{\nu}_{1}](x,x')
 = -2ie^2a^{D-2}\beta \bar{\eta}^{\mu\nu}K\delta^D(x-x'),
\end{equation}
where $K$ is the constant part of $f(y)$,
\begin{equation}
    K=\frac{2^{\varepsilon}\Gamma(3-\varepsilon)}
           {4\Gamma^2(2-\frac{\varepsilon}{2})}
    \bigg[\frac{1}{{\tt s}}
         + \pi\cot\big(\frac{\pi\varepsilon}{2}\big)-\gamma_{\rm E}
         - \psi(3-\varepsilon)
    \bigg]=\frac{1}{\varepsilon}
    + \frac{1}{2{\tt s}}-\frac{5}{4}+\ln(2)+O(\varepsilon).
\end{equation}
We shall call the left over portion of (\ref{eq - app fig.2}),
$i[^{\mu}\Pi^{\nu}_{2}](x,x')$ and use a bar over the metric tensor to
indicate that its zeroth components have been removed,
%
$
    \bar{\eta}^{\mu\nu}=\eta^{\mu\nu}+\delta^{\mu}_0\delta^{\nu}_0
$
.
%
Calculating the derivatives of $f(y)$ and dropping some unimportant
$O(\varepsilon)$ terms one easily finds
\begin{equation}
    f'^2-ff''=-\frac{1}{1-\frac{\varepsilon}{2}}\frac{1}{y^{4-\varepsilon}}
    +\frac{2-\frac{\varepsilon}{2}}{\varepsilon}\frac{1}{y^{3-\varepsilon}}
    -\frac{(2-\frac{\varepsilon}{2})K}{y^{3-\frac{\varepsilon}{2}}}
    +\frac{1}{y^2}\Big(\frac{1}{4}\ln\big(\frac{y}{4}\big)
    -\frac{1}{4{\tt s}}+\frac{3}{4}\Big)
\,,
\label{eq - derivatives func start}
\end{equation}
and
\begin{equation}
    ff'=-\frac{1}{1-\frac{\varepsilon}{2}}\frac{1}{y^{3-\varepsilon}}
    + \frac{(1-\varepsilon)(2-\frac{\varepsilon}{2})}
          {2\varepsilon(1-\frac{\varepsilon}{2})}\frac{1}{y^{2-\varepsilon}}
    - \frac{K}{y^{2-\frac{\varepsilon}{2}}}
    + \frac{1}{y}\Big(\frac{1}{4}\ln\big(\frac{y}{4}\big)
    - \frac{1}{4{\tt s}}+\frac{1}{2}\Big).
\end{equation}
Differentiating (\ref{eq - def g}) and keeping only terms up to
order $\varepsilon^0$ yields
\begin{equation}
    g'(y)=\frac{3}{y}\mspace{44mu},\mspace{44mu}g''(y)=-\frac{3}{y^2}.
\end{equation}
When we perform the summation in (\ref{eq - def h}) and calculate
the derivatives of the resulting expression, we obtain
\begin{align}
    h'(y)&=\frac{y-4+2(y-6)\ln(\frac{y}{4})}{(y-4)^2},\\
    h''(y)&=\frac{48-16y+16y\ln(\frac{y}{4})+y^2-2y^2\ln(\frac{y}{4})}{y(y-4)^3}.
    \label{eq - derivatives func stop}
\end{align}
Upon inserting~(\ref{eq - derivatives y start}) into~(\ref{eq -
derivatives y stop}) 
and~(\ref{eq - derivatives func start}--\ref{eq - derivatives func stop})
into (\ref{eq - app fig.2 bracket}), one gets
%
\begin{align}
    i[^{\mu}\Pi^{\nu}_{2}](x,x') 
 = & 2e^2a^{3-\varepsilon}a'^{3-\varepsilon}H^2\beta^2
\Bigg[
      4aa'H^2\Delta x^{\mu}\Delta x^{\nu}
\nonumber\\
    &\mspace{2mu}\times
   \bigg\{
      \frac{1}{1-\frac{\varepsilon}{2}}\frac{1}{y^{4-\varepsilon}}
    - \frac{2-\frac{\varepsilon}{2}}{\varepsilon}\frac{1}{y^{3-\varepsilon}}
    + \frac{(2-\frac{\varepsilon}{2})K}{y^{3-\frac{\varepsilon}{2}}}
    - \frac{1}{y^2}\Big(\frac{1}{4}\ln\big(\frac{y}{4}\big)
                      - \frac{1}{4{\tt s}}+\frac{3}{4}
                   \Big)
  \bigg\}
\nonumber\\
    & - 2\eta^{\mu\nu}
   \bigg\{\frac{1}{1-\frac{\varepsilon}{2}}\frac{1}{y^{3-\varepsilon}}
    -\frac{(1-\varepsilon)(2-\frac{\varepsilon}{2})}
          {2\varepsilon(1-\frac{\varepsilon}{2})}\frac{1}{y^{2-\varepsilon}}
    +\frac{K}{y^{2-\frac{\varepsilon}{2}}}
    -\frac{1}{y}\Big(\frac{1}{4}\ln\big(\frac{y}{4}\big)
                  -  \frac{1}{4{\tt s}}+\frac{1}{2}
                \Big)
   \bigg\}
\nonumber\\
    &
  +\bigg\{\frac{1}{y^2}\Big(\frac{1}{2}\ln{\big(\frac{y}{4}\big)}
     -\frac{1}{2{\tt s}}+2\Big)+\frac{1}{4y}
   \bigg\}
\Big(y\delta^{\mu}_0\delta^{\nu}_0
   -2a'H\Delta x^{\mu}\delta^{\nu}_0+2aH\Delta x^{\nu}\delta^{\mu}_0\Big)\Bigg]
\nonumber\\
    &\mspace{-30mu}-4e^2a^3a'^3\frac{H^6}{(4\pi)^4}
\Bigg[4aa'H^2\Delta x^{\mu}\Delta x^{\nu}
    \bigg\{\frac{3}{y^2}-\frac{48-16y+16y\ln\big(\frac{y}{4}\big)
           + y^2-2y^2\ln\big(\frac{y}{4}\big)}{y(y-4)^3}
    \bigg\}
\nonumber\\
    &-2\eta^{\mu\nu}
     \bigg\{
            \frac{3}{y}+\frac{y-4+2(y-6)\ln\big(\frac{y}{4}\big)}{(y-4)^2}
     \bigg\}
\nonumber\\
    &+\frac{2(y-4)(y-8)+4(12-y)\ln(\frac{y}{4})}{(y-4)^3}\Big(y\delta^{\mu}_0\delta^{\nu}_0
    -2a'H\Delta x^{\mu}\delta^{\nu}_0+2aH\Delta x^{\nu}\delta^{\mu}_0\Big)\Bigg]
\,,
    \label{eq - ipi2}
\end{align}
where here and in the rest of the appendices upper indices are raised by
the Minkowski metric $\eta^{\mu\nu}$. 
The portion of (\ref{eq - ipi2}) within the first bracket and the
constant in front of it can be written as
\begin{align}
    \Rightarrow&-\frac{4e^2\beta^2H^{2\varepsilon-4}}{1-\frac{\varepsilon}{2}}
    \Big[\eta^{\mu\nu}-2\frac{\Delta x^{\mu}\Delta x^{\nu}}{\Delta x^2}
    \Big]\frac{1}{\Delta x^{6-2\varepsilon}}
\nonumber\\
    &+2e^2\beta^2H^{2\varepsilon-2}aa'
       \frac{2-\frac{\varepsilon}{2}}{\varepsilon(1-\frac{\varepsilon}{2})}
    \Big[(1-\varepsilon)\eta^{\mu\nu}-(4-2\varepsilon)
                 \frac{\Delta x^{\mu}\Delta x^{\nu}}{\Delta x^2}
    \Big]\frac{1}{\Delta x^{4-2\varepsilon}}
\nonumber\\
    &-4e^2\beta^2H^{\varepsilon-2}a^{1-\frac{\varepsilon}{2}}a'^{1-\frac{\varepsilon}{2}}K
    \Big[\eta^{\mu\nu}-(4-\varepsilon)\frac{\Delta x^{\mu}\Delta x^{\nu}}{\Delta x^2}
    \Big]\frac{1}{\Delta x^{4-\varepsilon}}\nonumber\\
    &+2e^2\beta^2a^2a'^2\Big[\big\{-\frac{1}{Ha}\Delta x^{\mu}\delta^{\nu}_0
    +\frac{1}{Ha'}\Delta x^{\nu}\delta^{\mu}_0\big\}
        \big(\ln(\frac{y}{4})
            -\frac{1}{{\tt s}}+4
         \big)
   -\Delta x^{\mu}\Delta x^{\nu}
         \big(\ln(\frac{y}{4})-\frac{1}{{\tt s}}+3
         \big)\Big]\frac{1}{\Delta x^{4}}
    \nonumber\\
    &+2e^2\beta^2a^2a'^2\Big[
                             \bar{\eta}^{\mu\nu}\big\{\frac{1}{2}
                             \ln(\frac{y}{4})-\frac{1}{2{\tt s}}+1\big\}
                           + \delta^{\mu}_0\delta^{\nu}_0
                           - \frac{1}{2}a'H\Delta x^{\mu}\delta^{\nu}_0
                           + \frac{1}{2}aH\Delta x^{\nu}\delta^{\mu}_0
                        \Big]
    \frac{1}{\Delta x^{2}}\nonumber\\
    &+\frac{1}{2}e^2\beta^2a^3a'^3H^2\delta^{\mu}_0\delta^{\nu}_0.
\label{eq - 1st part vac pol}
\end{align}
We want to write the vacuum polarization in a form which makes
explicit that
$\partial_{\mu}i[^{\mu}\Pi^{\nu}](x,x')
= \partial_{\nu}'i[^{\mu}\Pi^{\nu}](x,x')=0$.
To find such a manifestly transverse form we will use the
following identities
\begin{equation}
    \Big[\eta^{\mu\nu}-2\frac{\Delta x^{\mu}\Delta x^{\nu}}{\Delta x^2}
    \Big]\frac{1}{\Delta x^{6-2\varepsilon}}=
    -\frac{1}{2(2-\varepsilon)(3-\varepsilon)}\big[\phantom{\,}^{\mu} P^{\nu}\big]
    \frac{1}{\Delta x^{4-2\varepsilon}},
\end{equation}
\begin{equation}
    \Big[(1-\varepsilon)\eta^{\mu\nu}-(4-2\varepsilon)\frac{\Delta x^{\mu}\Delta x^{\nu}}{\Delta x^2}
    \Big]\frac{1}{\Delta x^{4-2\varepsilon}}=
    -\frac{1}{2-2\varepsilon}\big[\phantom{\,}^{\mu} P^{\nu}\big]\frac{1}{\Delta x^{2-2\varepsilon}}
,
\end{equation}
\begin{equation}
    \Big[\eta^{\mu\nu}-(4-\varepsilon)\frac{\Delta x^{\mu}\Delta x^{\nu}}{\Delta x^2}
    \Big]\frac{1}{\Delta x^{4-\varepsilon}}=
    -\frac{1}{2-\varepsilon}\big[\phantom{\,}^{\mu} P^{\nu}\big]
    \frac{1}{\Delta x^{2-\varepsilon}}
    -\frac{2i\pi^{2-\frac{\varepsilon}{2}}}{\Gamma(2-\frac{\varepsilon}{2})}\bar{\eta}^{\mu\nu}\delta^{D}(x-x').
    \label{eq - transverse3}
\end{equation}
When inserting (\ref{eq - transverse3}) into (\ref{eq - 1st part vac pol}), 
the local term exactly cancels
$i[^{\mu}\Pi^{\nu}_{1}](x,x')$. We find for the remaining part of
(\ref{eq - 1st part vac pol}),
\begin{align}
\!\!\!\!\!\!
    \Rightarrow\frac{\alpha}{2\pi^3}
\Bigg[&\frac{\pi^{\varepsilon}\Gamma^2(1-\frac{\varepsilon}{2})}
            {2(3-\varepsilon)}
    \big[\phantom{\,}^{\mu} P^{\nu}\big]
    \frac{1}{\Delta x^{4-2\varepsilon}}
    -\frac{1}{\eta\eta'}\big[\phantom{\,}^{\mu} P^{\nu}\big]
    \frac{\frac{1}{2}\ln\big(\frac{y}{4}\big)
    -\frac{1}{2{\tt s}}+2}{\Delta x^{2}}
\nonumber\\
    &+\frac{1}{\eta^2\eta'^2}
  \bigg\{\big(\eta\Delta x^{\mu}\delta^{\nu}_0
    -\eta'\Delta x^{\nu}\delta^{\mu}_0-\Delta x^{\mu}\Delta x^{\nu}\big)
            \Big(\ln\big(\frac{y}{4}\big)-\frac{1}{{\tt s}}+3\Big)
\nonumber\\
    &\mspace{72mu}+(\eta-\eta')\big(\eta^{\mu\nu}(\eta-\eta')+\Delta x^{\mu}\delta^{\nu}_0+\Delta x^{\nu}\delta^{\mu}_0\big)
  \bigg\}\frac{1}{\Delta x^{4}}
\nonumber\\
    &+\frac{1}{\eta^2\eta'^2}
   \bigg\{\bar{\eta}^{\mu\nu}
          \Big(\frac{1}{2}\ln\big(\frac{y}{4}\big)-\frac{1}{2{\tt s}}+1\Big)
    +\delta^{\mu}_0\delta^{\nu}_0
     \!+\! \frac{1}{2\eta'}\Delta x^{\mu}\delta^{\nu}_0
    \!-\! \frac{1}{2\eta}\Delta x^{\nu}\delta^{\mu}_0
   \bigg\}
    \frac{1}{\Delta x^{2}}
    +\frac{\delta^{\mu}_0\delta^{\nu}_0}{4\eta^3\eta'^3}\Bigg]
,
\end{align}
where $\alpha = e^2/4\pi$ is the fine structure constant, and 
we dropped all $O(\varepsilon)$ terms, except in the first
line. These terms are not required for regularization.  Note that
the second term is not yet in manifestly transverse form, because
of the factor $\frac{1}{\eta\eta'}$ standing to the left of the
projector operator. Bringing it to the right and combining with
the rest of (\ref{eq - ipi2}) we get for the sum of all terms 
in Eqs.~(\ref{eq - app fig.1}) and~(\ref{eq - app fig.2}) 
%
\begin{align}
i&[^{\mu}\Pi^{\nu}_{1+2}](x,x')
  = \frac{\alpha}{2\pi^3}
\Bigg[
    \frac{\pi^{\varepsilon}\Gamma^2(1-\frac{\varepsilon}{2})}{2(3-\varepsilon)}
    \big[\phantom{\,}^{\mu} P^{\nu}\big]
    \frac{1}{\Delta x^{4-2\varepsilon}}
    -\big[\phantom{\,}^{\mu} P^{\nu}\big]
    \frac{\frac{1}{2}\ln(\frac{y}{4})
    -\frac{1}{2{\tt s}}+2}{\eta\eta'\Delta x^{2}}\nonumber\\
    &+\frac{1}{\eta^4\eta'^4 y^2}
    \bigg\{
           (\eta-\eta')\big(\eta^{\mu\nu}(\eta-\eta')
       +   \Delta x^{\mu}\delta^{\nu}_0+\Delta x^{\nu}\delta^{\mu}_0\big)
           \Big(\!\ln\big(\frac{y}{4}\big)\!-\!\frac{1}{{\tt s}}+4\Big)
       -   \Delta x^{\mu}\Delta x^{\nu}
           \Big(\!\ln\big(\frac{y}{4}\big)\!-\!\frac{1}{{\tt s}}+3\Big)
  \bigg\}
\nonumber\\
    &+\frac{1}{\eta^3\eta'^3 y}
   \bigg\{
       -  2\eta^{\mu\nu}-\delta^{\mu}_0\delta^{\nu}_0
       +  \frac{1}{2\eta'}\Delta x^{\mu}\delta^{\nu}_0
       -  \frac{1}{2\eta}\Delta x^{\nu}\delta^{\mu}_0
   \bigg\}
    +\frac{\delta^{\mu}_0\delta^{\nu}_0}{4\eta^3\eta'^3}\nonumber\\
    &-\frac{\Delta x^{\mu}\Delta x^{\nu}}{2\eta^4\eta'^4}
    \bigg\{
           \frac{3}{y^2}
        -  \frac{48-16y+16y\ln(\frac{y}{4})+y^2-2y^2\ln(\frac{y}{4})}
                {y(y-4)^3}
    \bigg\}
\nonumber\\
    &+\frac{\eta^{\mu\nu}}{4\eta^3\eta'^3}
  \bigg\{
         \frac{3}{y}+\frac{y-4+2(y-6)\ln(\frac{y}{4})}{(y-4)^2}
  \bigg\}
\nonumber\\
    &-\frac{1}{\eta^3\eta'^3}\frac{\frac{1}{4}(y-4)(y-8)+\frac{1}{2}(12-y)\ln(\frac{y}{4})}{(y-4)^3}
\Big(y\delta^{\mu}_0\delta^{\nu}_0
    +\frac{2}{\eta'}\Delta x^{\mu}\delta^{\nu}_0-\frac{2}{\eta}\Delta x^{\nu}\delta^{\mu}_0\Big)\Bigg].
\label{eq - ipi1+2}
\end{align}
We will use the following ansatz for the terms in (\ref{eq - ipi1+2}),
which are not in manifestly transverse form
\begin{align}
    \frac{\alpha}{2\pi^3}
\Big[\big[\phantom{\,}^{\mu} P^{\nu}\big]\frac{u(y)}{\eta^2\eta'^2}+
    \big[\phantom{\,}^\mu\bar P^\nu\big]&\frac{w(y)}{\eta^2\eta'^2}\Big]
    =\frac{\alpha}{2\pi^3}
\Bigg[\frac{\delta^{\mu}_0\delta^{\nu}_0}{\eta^3\eta'^3}
    \big\{-4u-5u'y-u''y^2-4w'-4w''y\big\}\nonumber\\
    &+\frac{\eta^{\mu\nu}}{\eta^3\eta'^3}
      \big\{-4u+u'(6-5y)-u''y^2-4w'+4w''(2-y)\big\}
\nonumber\\
    &+\frac{\Delta x^{\mu}\Delta x^{\nu}}{\eta^4\eta'^4}\big\{4u''+4w''\big\}
    +\Big\{\frac{\Delta x^{\nu}\delta^{\mu}_0}{\eta^4\eta'^3}
    -\frac{\Delta x^{\mu}\delta^{\nu}_0}{\eta^3\eta'^4}\Big\}4w''
    \nonumber\\
    &+\Big\{\frac{\Delta x^{\mu}\delta^{\nu}_0}{\eta^4\eta'^3}
         -  \frac{\Delta x^{\nu}\delta^{\mu}_0}{\eta^3\eta'^4}
         -  \frac{\eta^{\mu\nu}}{\eta^4\eta'^2}
         -  \frac{\eta^{\mu\nu}}{\eta^2\eta'^4}
      \Big\}\big\{6u'+2u''y+4w''\big\}
\Bigg]
\,.
\label{eq - ansatz transverse}
\end{align}
Comparing this to the relevant terms in~(\ref{eq - ipi1+2}), one
finds
\begin{equation}
    4w''=-\frac{1}{y^2}\Big(\ln\big(\frac{y}{4}\big)
                          - \frac{1}{{\tt s}}+4
                       \Big)-\frac{1}{2y}
    +2\frac{\frac{1}{4}(y-4)(y-8)+\frac{1}{2}(12-y)\ln(\frac{y}{4})}{(y-4)^3},
\end{equation}
\begin{equation}
    6u'+2u''y+4w''=-\frac{1}{y^2}\Big(\ln\big(\frac{y}{4}\big)
                                    - \frac{1}{{\tt s}}+4
                                 \Big)
\,,
\end{equation}
\begin{equation}
 4u''+4w''
   = -\frac{1}{y^2}\Big(
                        \ln\big(\frac{y}{4}\big)
                      - \frac{1}{{\tt s}}+3
                   \Big)
     -\frac{1}{2}
       \Big\{\frac{3}{y^2}-\frac{48-16y+16y
               \ln(\frac{y}{4})+y^2-2y^2\ln(\frac{y}{4})}{y(y-4)^3}
       \Big\}
,
\end{equation}
\begin{align}
    4u-u'(6-5y)+u''y^2+4w'-&4w''(2-y)=\frac{2}{y^2}\big(\ln(\frac{y}{4})
      -\frac{1}{{\tt s}}+4\big)+\frac{2}{y}
\nonumber\\
    &
  -\frac{1}{4}\Big\{\frac{3}{y}+\frac{y-4+2(y-6)\ln(\frac{y}{4})}{(y-4)^2}
              \Big\}
,
\end{align}
\begin{align}
    4u+5u'y+u''y^2+4w'+4w''y
 = \frac{1}{y}
 - \frac{1}{4}+y\frac{\frac{1}{4}(y-4)(y-8)
 + \frac{1}{2}(12-y)\ln(\frac{y}{4})}{(y-4)^3}.
\end{align}
One can find the following solutions to these differential
equations
\begin{align}
    u(y)=&-\frac{\ln(\frac{y}{4})}{2(y-4)},\\
    w(y)=&\frac{1}{8}\ln^2\big(\frac{y}{4}\big)
        + \ln\big(\frac{y}{4}\big)
               \Big(1-\frac{1}{4{\tt s}}+\frac{y-2}{2(y-4)}\Big)
                - \frac{1}{4}Li_2\big(1-\frac{y}{4}\big),
\end{align}
where the dilogarithm function, defined by
$Li_2(z)\equiv-\int_0^{z}\frac{\ln(1-t)}{t}dt$ appears. Upon inserting
these solutions into (\ref{eq - ansatz transverse}) and combining
the result with the first two terms of (\ref{eq - ipi1+2}), one
obtains the following expression for the graphs in Fig.\ref{fig - vac
pol graphs}.(1) and Fig.\ref{fig - vac pol graphs}.(2)
\begin{align}
    i[^{\mu}\Pi^{\nu}_{1+2}](x,x')&=\frac{\alpha}{2\pi^3}
\Bigg[
    \big[\phantom{\,}^{\mu} P^{\nu}\big]
 \bigg\{\frac{\pi^{\varepsilon}\Gamma^2
       (1-\frac{\varepsilon}{2})}{2(3-\varepsilon)}
    \frac{1}{\Delta x^{4-2\varepsilon}}
      -\frac{\frac{1}{2}(\ln(\frac{y}{4})
      -\frac{1}{{\tt s}}+2)+1}{\eta\eta'\Delta x^{2}}
    -\frac{\ln\big(\frac{y}{4}\big)}{2(y-4)\eta^2\eta'^2}
  \bigg\}
\nonumber\\
    &+\big[\phantom{\,}^\mu\bar P^\nu\big]
    \frac{\frac{1}{8}\ln^2\big(\frac{y}{4}\big)
    + \ln\big(\frac{y}{4}\big)\big(1-\frac{1}{4{\tt s}}+\frac{y-2}{2(y-4)}\big)
    -\frac{1}{4}Li_2\big(1-\frac{y}{4}\big)}{\eta^2\eta'^2}+O({\tt s})
\Bigg]
\,.
\end{align}

\section{Integration of the retarded vacuum polarization against
         the photon field}
 \label{sec - integral in field equation}

In order to simplify the effective field equation for the photon
field, we need to evaluate the integral
\begin{equation}
    \int d^4x'[^{\mu}\Pi^{r,\nu}_{ren}](x,x')A_\nu(x'),
    \label{eq - integral}
\end{equation}
where
\begin{equation}
    [^{\mu}\Pi^{r,\nu}_{ren}](x,x')=\frac{\alpha}{8\pi^2{\tt s}}
              \Big[\big[\phantom{\,}^{\mu} P^{\nu}\big]
    \frac{\partial^2\Theta(|\Delta\eta|-\|\Delta\vec{x}\|)\Theta(\Delta\eta)}{\eta\eta'}
    -\big[\phantom{\,}^\mu\bar P^\nu\big]
    \frac{2\Theta(|\Delta\eta|-\|\Delta\vec{x}\|)\Theta(\Delta\eta)}{\eta^2\eta'^2}\Big].
\end{equation}
and
\begin{equation}
    A_{\nu}(x')=\varepsilon_{\nu}(\vec{k},\eta')e^{i\vec{k}\cdot\vec{x}'},
\end{equation}
\begin{equation}
    \Big(\partial'_0-\frac{2}{\eta'}\Big)
       \varepsilon_0(\vec{k},\eta')=i\vec{k}\cdot\vec\varepsilon(\vec{k},\eta')
\,.
    \label{eq - app eps gauge}
\end{equation}
We can simplify the integral by using the identity,
\begin{align}
    \big[\phantom{\,}^{\mu} P^{\nu}\big]&\frac{f(x-x')}{\eta\eta'}=
    \Big\{\bar{\eta}^{\mu\nu}\big[-\frac{1}{\eta\eta'}\partial^2+\frac{\Delta\eta}{\eta^2\eta'^2}\partial_0
    -\frac{1}{\eta^2\eta'^2}\big]+\frac{\delta^{\mu}_0\delta^{\nu}_0}{\eta\eta'}\vec{\nabla}^2\nonumber\\
    &\quad-\delta^{\nu}_0\bar{\partial}^{\mu}\big[\frac{1}{\eta\eta'}\partial_0-\frac{1}{\eta^2\eta'}\big]
    +\big[\frac{1}{\eta\eta'}\bar{\partial}^{\mu}-\frac{\delta^{\mu}_0}{\eta\eta'}\partial_0
    -\frac{\delta^{\mu}_0}{\eta\eta'^2}\big]\bar{\partial}^{\nu}\Big\}f(x-x').
\end{align}
where $f$ can be an arbitrary function of $x-x'$ and a bar over a
tensor is used to indicate that its zero components have been
removed, {\it e.g.}
$\bar{\eta}^{\mu\nu}\equiv\eta^{\mu\nu}+\delta^{\mu}_0\delta^{\nu}_0$,
$\bar{\partial}^{\mu}\equiv\partial^{\mu}-\delta^{\mu}_0\partial^0$.
Using this we find for the integral (\ref{eq - integral})
\begin{align}
    \Rightarrow\frac{\alpha}{8\pi^2{\tt s}}\int d^{4}x'\Bigg[
    \Big\{&\bar{\eta}^{\mu\nu}\big[-\frac{1}{\eta\eta'}\partial^2+\frac{\Delta\eta}{\eta^2\eta'^2}\partial_0
    -\frac{1}{\eta^2\eta'^2}\big]+\frac{\delta^{\mu}_0\delta^{\nu}_0}{\eta\eta'}\vec{\nabla}^2
    -\delta^{\nu}_0\bar{\partial}^{\mu}\big[\frac{1}{\eta\eta'}\partial_0-\frac{1}{\eta^2\eta'}\big]\nonumber\\
    &\mspace{40mu}+\big[\frac{1}{\eta\eta'}\bar{\partial}^{\mu}-\frac{\delta^{\mu}_0}{\eta\eta'}\partial_0
    -\frac{\delta^{\mu}_0}{\eta\eta'^2}\big]\bar{\partial}^{\nu}\Big\}
    \partial^2\Theta(|\Delta\eta|-\|\Delta\vec{x}\|)\Theta(\Delta\eta)\nonumber\\
    -&\frac{2}{\eta^2\eta'^2}\big[\phantom{\,}^\mu\bar P^\nu\big]
    \Theta(|\Delta\eta|-\|\Delta\vec{x}\|)\Theta(\Delta\eta)\Bigg]
    \varepsilon_{\nu}(\vec{k},\eta')e^{i\vec{k}\cdot\vec{x}'}.
    \label{eq - integral eom1}
\end{align}
In the last term the transverse projector
$\big[\phantom{\,}^\mu\bar P^\nu\big]$ hits a function of $x-x'$
only. Therefore it can be replaced by
$-(\bar{\eta}^{\mu\nu}\vec{\nabla}^2-\bar{\partial}^{\mu}\bar{\partial}^{\nu})$.
Then this expression can be recast as
\begin{align}
    \frac{\alpha}{8\pi^2{\tt s}}
     \Bigg[&-\frac{\bar{\eta}^{\mu\nu}}{\eta}\partial^4I_{1,\nu}(x)
    +\frac{\bar{\eta}^{\mu\nu}}{\eta}\partial_{0}\partial^{2}I_{2,\nu}(x)
    -\frac{\bar{\eta}^{\mu\nu}}{\eta^2}\partial_{0}\partial^{2}I_{1,\nu}(x)
    -\frac{\bar{\eta}^{\mu\nu}}{\eta^2}\partial^{2}I_{2,\nu}(x)\nonumber\\
    &+\frac{\delta^{\mu}_0\delta^{\nu}_0}{\eta}\vec{\nabla}^2\partial^{2}I_{1,\nu}(x)
    -\frac{\delta^\nu_0}{\eta}\bar{\partial}^\mu\partial_0\partial^{2}I_{1,\nu}(x)
    +\frac{\delta^\nu_0}{\eta^2}\bar{\partial}^\mu\partial^{2}I_{1,\nu}(x)\nonumber\\
    &+\frac{1}{\eta}\bar{\partial}^{\mu}\bar{\partial}^{\nu}\partial^{2}I_{1,\nu}(x)
    -\frac{\delta^{\mu}_0}{\eta}\partial_{0}\bar{\partial}^{\nu}\partial^{2}I_{1,\nu}(x)
    -\frac{\delta^{\mu}_0}{\eta}\bar{\partial}^{\nu}\partial^{2}I_{2,\nu}(x)\nonumber\\
    &+\frac{2\bar{\eta}^{\mu\nu}}{\eta^2}\vec{\nabla}^2I_{2,\nu}(x)-\frac{2}{\eta^2}\bar{\partial}^{\mu}\bar{\partial}^{\nu}I_{2,\nu}(x)\Bigg],
    \label{eq - integral eom2}
\end{align}
where
\begin{equation}
    I_{(1,2),\nu}(x)=\int d^4x'\frac{1}{\eta'^{(1,2)}}\Theta(|\Delta\eta|-\|\Delta\vec{x}\|)\Theta(\Delta\eta)
    \varepsilon_{\nu}(\vec{k},\eta')e^{i\vec{k}\cdot\vec{x}'}.
    \label{eq - def integral1,2}
\end{equation}
One can get rid of the step functions by choosing appropriate
limits of the integration. Then one can rewrite (\ref{eq - def
integral1,2}) as
\begin{equation}
    =e^{i\vec{k}\cdot\vec{x}}\int_{-\infty}^{\eta}d\eta'\frac{4\pi\varepsilon_{\nu}(\vec{k},\eta')}{\eta'^{(1,2)}}
    \int_0^{\Delta \eta}r^2dr
    \int_{-1}^{+1}\frac{d \cos{\theta}}{2}\thinspace
    e^{-ikr\cos\theta},
\end{equation}
where $\theta$ is the angle between $\Delta\vec{x}$ and $\vec{k}$,
$r=\|\Delta\vec{x}\|$ and $k=\|\vec{k}\|$. Performing the $r$ and
$\cos{\theta}$ integration yields
\begin{equation}
    =e^{i\vec{k}\cdot\vec{x}}\int_{-\infty}^{\eta}d\eta'\frac{\varepsilon_{\nu}(\vec{k},\eta')}{\eta'^{(1,2)}}
    \frac{4\pi}{k^3}\big(\sin(k\Delta\eta)-(k\Delta\eta)\cos(k\Delta\eta)\big).
    \label{eq - integral(1,2)}
\end{equation}
 From this expression we obtain
\begin{equation}
    \partial^2I_{(1,2),\nu}(x)=-e^{i\vec{k}\cdot\vec{x}}
    \int_{-\infty}^{\eta}d\eta'\frac{\varepsilon_{\nu}(\vec{k},\eta')}{\eta'^{(1,2)}}
    \frac{8\pi}{k}\sin(k\Delta\eta).
    \label{eq - partial2 integral(1,2)}
\end{equation}
Using (\ref{eq - partial2 integral(1,2)}) one can write the
contributions to the $\mu=0$ component of Eq. (\ref{eq - integral
eom2}) as
\begin{align}
    \Rightarrow\frac{\alpha}{\pi{\tt s}}\frac{\delta^\mu_0}{\eta}
\bigg[&k^2
    e^{i\vec{k}\cdot\vec{x}}
     \int_{-\infty}^{\eta}d\eta'\frac{\varepsilon_{0}(\vec{k},\eta')}{\eta'}
    \frac{1}{k}\sin(k\Delta\eta)\nonumber\\
    +&i\bar{k}^\nu\partial_0e^{i\vec{k}\cdot\vec{x}}
    \int_{-\infty}^{\eta}d\eta'\frac{\bar{\varepsilon}_{\nu}(\vec{k},\eta')}{\eta'}
    \frac{1}{k}\sin(k\Delta\eta)\nonumber\\
    +&i\bar{k}^\nu e^{i\vec{k}\cdot\vec{x}}
    \int_{-\infty}^{\eta}d\eta'\frac{\bar{\varepsilon}_{\nu}(\vec{k},\eta')}{\eta'^2}
    \frac{1}{k}\sin(k\Delta\eta)
\bigg].
\end{align}
Upon using (\ref{eq - app eps gauge}) and performing partial
integration this can be recast as
\begin{align}
    \Rightarrow\frac{\alpha}{\pi{\tt s}}\frac{\delta^\mu_0}{\eta}
\Bigg[&k^2
 e^{i\vec{k}\cdot\vec{x}}\int_{-\infty}^{\eta}
         d\eta'\varepsilon_{0}(\vec{k},\eta')
    \frac{1}{k}\frac{\sin(k\Delta\eta)}{\eta'}
\nonumber\\
  +&e^{i\vec{k}\cdot\vec{x}}\partial_0
\bigg\{\int_{-\infty}^{\eta}d\eta'\varepsilon_{0}(\vec{k},\eta')
    \frac{1}{k}
\Big(\frac{k\cos(k\Delta\eta)}{\eta'}+\frac{\sin(k\Delta\eta)}{\eta'^2}\Big)
-2\int_{-\infty}^{\eta}d\eta'\varepsilon_{0}(\vec{k},\eta')
    \frac{1}{k}\frac{\sin(k\Delta\eta)}{\eta'^2}
\bigg\}
\nonumber\\
    +&e^{i\vec{k}\cdot\vec{x}}
\bigg\{\int_{-\infty}^{\eta}d\eta'\varepsilon_{0}(\vec{k},\eta')
    \frac{1}{k}\Big(\frac{k\cos(k\Delta\eta)}{\eta'^2}
    +\frac{2\sin(k\Delta\eta)}{\eta'^3}\Big)
-2\int_{-\infty}^{\eta}d\eta'\varepsilon_{0}(\vec{k},\eta')
    \frac{1}{k}\frac{\sin(k\Delta\eta)}{\eta'^3}
\bigg\}
\Bigg].
\nonumber
\end{align}
When performing the $\partial_0$ differentiation all integrals
cancel, only one term remains. It comes from the differentiation
with respect to the upper limit of the integral $
\int_{-\infty}^{\eta}d\eta'\varepsilon_{0}(\vec{k},\eta')
\frac{1}{k}\frac{k\cos(k\Delta\eta)}{\eta'}$, thus we get
\begin{equation}
    \Rightarrow-\frac{\alpha}{\pi{\tt s}}
            \frac{\delta^\mu_0}
                 {\eta^2}\varepsilon^{0}(\vec{k},\eta)e^{i\vec{k}\cdot\vec{x}}
\,.
    \label{eq - integral mu=0}
\end{equation}
The terms contributing to the spatial components of (\ref{eq -
integral eom2}) can be written as
\begin{align}
    \frac{\alpha}{\pi{\tt s}}
     \Bigg[&-\frac{\bar{\eta}^{\mu\nu}}{\eta}(\partial_0^2+k^2)
    e^{i\vec{k}\cdot\vec{x}}\int_{-\infty}^{\eta}d\eta'\frac{\varepsilon_{\nu}(\vec{k},\eta')}{\eta'}
    \frac{1}{k}\sin(k\Delta\eta)\nonumber\\
    &-\frac{\bar{\eta}^{\mu\nu}}{\eta}\partial_0
    e^{i\vec{k}\cdot\vec{x}}\int_{-\infty}^{\eta}d\eta'\frac{\varepsilon_{\nu}(\vec{k},\eta')}{\eta'^2}
    \frac{1}{k}\sin(k\Delta\eta)\nonumber\\
    &+\frac{\bar{\eta}^{\mu\nu}}{\eta^2}\partial_0
    e^{i\vec{k}\cdot\vec{x}}\int_{-\infty}^{\eta}d\eta'\frac{\varepsilon_{\nu}(\vec{k},\eta')}{\eta'}
    \frac{1}{k}\sin(k\Delta\eta)\nonumber\\
    &+\frac{\bar{\eta}^{\mu\nu}}{\eta^2}
    e^{i\vec{k}\cdot\vec{x}}\int_{-\infty}^{\eta}d\eta'\frac{\varepsilon_{\nu}(\vec{k},\eta')}{\eta'^2}
    \frac{1}{k}\sin(k\Delta\eta)\nonumber\\
    &+\frac{1}{\eta}(i\bar{k}^\mu)\partial_0 e^{i\vec{k}\cdot\vec{x}}\int_{-\infty}^{\eta}d\eta'\frac{\varepsilon_{0}(\vec{k},\eta')}{\eta'}
    \frac{1}{k}\sin(k\Delta\eta)\nonumber\\
    &-\frac{1}{\eta^2}(i\bar{k}^\mu)e^{i\vec{k}\cdot\vec{x}}\int_{-\infty}^{\eta}d\eta'\frac{\varepsilon_{0}(\vec{k},\eta')}{\eta'}
    \frac{1}{k}\sin(k\Delta\eta)\nonumber\\
    &-\frac{1}{\eta}(i\bar{k}^\mu)(i\bar{k}^\nu)e^{i\vec{k}\cdot\vec{x}}\int_{-\infty}^{\eta}d\eta'\frac{\varepsilon_{\nu}(\vec{k},\eta')}{\eta'}
    \frac{1}{k}\sin(k\Delta\eta)\nonumber\\
    &-\frac{\bar{\eta}^{\mu\nu}}{\eta^2}
    e^{i\vec{k}\cdot\vec{x}}\int_{-\infty}^{\eta}d\eta'\frac{\varepsilon_{\nu}(\vec{k},\eta')}{\eta'^{2}}
    \frac{1}{k}\big(\sin(k\Delta\eta)-(k\Delta\eta)\cos(k\Delta\eta)\big)\nonumber\\
    &-\frac{1}{\eta^2}(i\bar{k}^\mu)(i\bar{k}^\nu)e^{i\vec{k}\cdot\vec{x}}\int_{-\infty}^{\eta}d\eta'\frac{\varepsilon_{\nu}(\vec{k},\eta')}{\eta'^{2}}
    \frac{1}{k^3}\big(\sin(k\Delta\eta)-(k\Delta\eta)\cos(k\Delta\eta)\big)\Bigg],
\end{align}
where (\ref{eq - integral(1,2)}) and (\ref{eq - partial2
integral(1,2)}) have been used. We now insert Eq.~(\ref{eq - app eps gauge})
into the terms in which 
$i\bar{k}^\nu\varepsilon_\nu(\vec{k},\eta')$ appears and partially
integrate. When we then perform the remaining differentiations all
integrals cancel, and only a term, which comes from the
differentiation with respect to the upper limit of the first
integral, remains. This term equals
\begin{equation}
    \Rightarrow-\frac{\alpha}{\pi{\tt s}}
            \frac{1}{\eta^2}
            \bar{\varepsilon}^{\mu}(\vec{k},\eta)e^{i\vec{k}\cdot\vec{x}}.
\end{equation}
Combining this with (\ref{eq - integral mu=0}) we finally find for
the integral (\ref{eq - integral})
\begin{equation}
    \int d^4x'[^{\mu}\Pi^{r,\nu}_{ren}](x,x')A_\nu(x')
=
    -\frac{\alpha H^2}{\pi {\tt s}}a^2 \eta^{\mu\nu}A_{\nu}(x).
\end{equation}

\end{document}